\documentclass[10pt,a4paper]{article}

\usepackage{jheppub}

\usepackage{epsfig,epsf}
\usepackage{amsmath}
\usepackage{amsthm}
\usepackage{amsfonts}
\usepackage{amssymb}
\usepackage{dsfont}
\usepackage{upref}
\usepackage{amsopn}
\usepackage{bm}

\usepackage{color}

\usepackage[utf8]{inputenc}

\usepackage{epstopdf}

\usepackage{multirow}

\usepackage{nccmath}
\usepackage{float}

\usepackage{rotating}

\usepackage{marvosym}

\usepackage{subcaption}
\usepackage{array}   
\newcolumntype{C}{>{$}c<{$}}

\usepackage{slashed}

\usepackage[active]{srcltx}

\newcommand \widebar [1] {\overline{#1}}

\def\II{\hbox{{1}\kern-.25em\hbox{l}}}

\DeclareMathOperator{\Li}{Li}

\newcommand \vev [1] {\langle{#1}\rangle}
\newcommand \VEV [1] {\left\langle{#1}\right\rangle}

\newcommand \Dl {\stackrel{\leftarrow}{D}}
\newcommand \Dr {\stackrel{\rightarrow}{D}}

\def\II{\hbox{{1}\kern-.25em\hbox{l}}}
\def \Li {\text{Li\,}}

\title{
{\small\today}
\begin{flushright}
{\large \textnormal{DESY 18-202}}\\[2mm]
\end{flushright}
Two-loop evolution equations for flavor-singlet light-ray operators}

\author[a]{V. M. Braun,}
\author[b,a,c]{A. N. Manashov,}
\author[b]{S. Moch}
\author[a]{and M. Strohmaier}

\affiliation[a]{
   Institut f\"ur Theoretische Physik, Universit\"at
   Regensburg \\ D-93040 Regensburg, Germany}
\affiliation[b]{
   Institut f\"ur Theoretische Physik, Universit\"at Hamburg\\
   D-22761 Hamburg, Germany}
\affiliation[c]{
   St.Petersburg Department of Steklov
Mathematical Institute, 191023 St.Petersburg, Russia}

\emailAdd{vladimir.braun@ur.de}

\emailAdd{alexander.manashov@desy.de}

\emailAdd{sven-olaf.moch@desy.de}

\emailAdd{matthias.strohmaier@ur.de}

\abstract{
QCD in non-integer $d=4-2\epsilon$  space-time dimensions enjoys conformal invariance at
the special fine-tuned value of the coupling.
Counterterms for composite operators in minimal subtraction schemes do not depend on $\epsilon$ by construction,
and therefore the renormalization group equations for composite operators in physical (integer) dimensions inherit
conformal symmetry. This observation can be used to restore the complete evolution kernels that take into account
mixing with the operators containing total derivatives from their eigenvalues (anomalous dimensions).
Using this approach we calculate the two-loop (NLO) evolution kernels for the leading twist flavor-singlet
operators in the position space (light-ray operator) representation. As the main result of phenomenological
relevance, in this way we are able to confirm the evolution equations of flavor-singlet
generalized hadron parton distributions derived earlier by Belitsky and M{\"u}ller
{ using a different approach.}
       }

\keywords{Conformal symmetry, Perturbative QCD, NLO Computations}

\begin{document}
\maketitle

\section{Introduction}\label{sect:intro}

{ The exciting possibility to study the three-dimensional proton structure (``tomographic imaging'') at the next generation
lepton-hadron and lepton-nucleus colliders, most notably the EIC~\cite{Accardi:2012qut}, poses formidable theoretical challenges.
To meet these challenges precise predictions in quantum chromodynamics (QCD) are necessary and require pushing the corresponding
computational tools to the highest possible accuracy. In practice, however, there exists still a considerable gap between precision
and theoretical rigor applied in data analyses at the energy frontier, e.g., by the LHC community searching for New Physics beyond
the Standard Model, and the QCD studies at medium energy machines, where the main goal is an improved understanding of the strong
interactions.

This concerns, in particular, differences in the accuracy when considering the scale dependence of parton distributions.
Whereas the next-to-next-to-leading order (NNLO), i.e., three-loop, analysis of parton distributions and fragmentation functions
is becoming standard~\cite{Accardi:2016ndt}, the analysis of deeply-virtual Compton scattering and, e.g.,
hard exclusive vector meson electro-production is still often based on the leading-order (LO) evolution,
despite the complete next-to-leading order (NLO) evolution kernels being available for a long time~\cite{Belitsky:1999hf}.
The necessity to close this gap is becoming gradually accepted so that the task to verify the results
of Ref.~\cite{Belitsky:1999hf} by an independent calculation and to develop the techniques to push such calculations to
NNLO accuracy are high on the agenda.}

On the technical level, the challenge is that considering off-forward matrix elements one has to take into account mixing with the
operators containing total derivatives. The direct calculation of the relevant Feynman diagrams is difficult as they involve two
different momenta, so that the method of choice has been to use the constraints from conformal symmetry that allow one to ``save''
one loop in comparison to the direct computation. This approach was pioneered by Dieter M\"uller
\cite{Mueller:1991gd,Mueller:1993hg,Mueller:1997ak} and further refined in
Refs.~\cite{Belitsky:1998gc,Belitsky:1998vj,Belitsky:1999hf}. The existing results for the two-loop (NLO) evolution kernels
\cite{Belitsky:1999hf} have been obtained using this technique.

In Ref.~\cite{Braun:2013tva} we proposed a somewhat different implementation of the same idea, based on the exact conformal
symmetry  of QCD in $d=4-2\epsilon$ dimensions at the critical point. Another difference to M\"uller's approach is our use of the
position space (light-ray operator) representation
{%
that allows one to switch rather easily to momentum fraction space, }
avoiding the generally nontrivial problem of the restoration of the evolution kernels from the mixing matrices of local operators.
Evolution kernels in position space are also interesting on their own in connection with lattice QCD calculations of Euclidean
``observables'' that can be factorized in terms of parton distributions, see e.g.,
{%
Refs.}~\cite{Braun:2007wv,Ji:2013dva,Ma:2017pxb}.
Such calculations currently attract a lot of attention.

This modified approach was tested in {
~\cite{Braun:2013tva} on several examples to two- and three-loop accuracy for scalar theories. In
{
~\cite{Braun:2014vba, Braun:2016qlg,Braun:2017cih} we used it to calculate the NLO
(two-loop)~\cite{Braun:2014vba} and NNLO (three-loop)~\cite{Braun:2016qlg,Braun:2017cih} evolution kernels for the leading-twist
quark-antiquark flavor-nonsinglet operators. In this work we address the flavor-singlet sector. This case is technically more
complicated and also new questions arise concerning the role of gauge-noninvariant operators in conformal Ward identities. These
issues will be discussed in what follows. Our main result is the derivation of the complete set of NLO evolution kernels in the
position space representation for the leading twist gluon and (C-even) quark-antiquark operators. Expanding these kernels at small
field separations we reproduce the results for the mixing matrices for flavor-singlet local operators given in
Ref.~\cite{Belitsky:1998gc}. We also demonstrate how the evolution kernels in momentum fraction space can be obtained from the
position space expressions by simple integration. We have verified numerically that the resulting expressions agree with the
kernels that are implemented
{%
in the NLO evolution code for generalized parton distributions (GPD)}
by Freund and McDermott~\cite{Freund:2001rk,Freund:2001hd} which is based on the analytic expressions derived in
Ref.~\cite{Belitsky:1999hf}.

\section{Flavor-singlet light-ray operators}\label{sect:operators}

The general formalism {%
of evolution equations for light-ray operators}
goes back to Ref.~\cite{Balitsky:1987bk} and is described in detail
in {
~\cite{Braun:2014vba,Braun:2016qlg,Braun:2017cih} so that we will be brief in what follows.
In this work we study the evolution equations for the flavor-singlet $C$-even twist-two light-ray operators (LROs) which are
defined as~\footnote{All equations are written assuming Euclidean space.}
\begin{subequations}
\label{Two-OP}
\begin{align}
\mathcal{O}_g(z_1,z_2)  &= F^a_{+\mu}(z_1n)[z_1,z_2]^{ab} F^b_{+\mu}(z_2 n)\,, \label{gluon-op}
\\
\label{quark-op}
\mathcal{O}_q(z_1,z_2)  &= \frac{1}2\Big( \bar q^{i,f}(z_1n) [z_1,z_2]^{ij}  \gamma_+\,q^{j,f}(z_2 n) -
(z_1\leftrightarrow z_2)\Big)\,.
\end{align}
\end{subequations}
Here $n^\mu$ is an auxiliary light-like vector, $n^2=0$, the ``plus'' projection stands for $F_{+\mu} = n^\nu F_{\nu\mu}$, and
$z_1,z_2$ are real numbers. We will often omit the factor $n^\mu$ in the argument specifying the field position and use a
short-hand notation $q(z) \equiv q(z n)$, etc. The gauge links in {%
Eq.}~\eqref{Two-OP} are taken in the adjoint and fundamental
representations, respectively,
\begin{align}
[z_1,z_2]=P\exp\left\{igz_{12} \int_0^1 du A_+(z_{21}^u)\right\}\,,
\label{link}
\end{align}
where $A_+=A_+^a T^a$ and in the adjoint representation  $T^a_{bb'}=if^{bab'}$.
In {%
Eq.~\eqref{link} we have} introduced another notation that will be used throughout this work:
\begin{align}
  z_{12} = z_1-z_2\,, \qquad z_{21}^u = \bar u z_2 + u z_1\,, \qquad \bar u =1-u\,.
\end{align}

The LROs~\eqref{Two-OP} can be viewed as the generating functions for local quark and gluon operators
\begin{align}\label{LRO-def-bare}
\mathcal{O}_\chi(z_1,z_2)=\sum\limits_{n=1}^\infty \sum\limits_{k=n}^\infty \Psi^{(\chi)}_{nk}(z_1,z_2)  \mathcal{O}_{\chi,nk}(0)\,,
\end{align}
where $\chi = q,g$ and one can choose, for example,
\begin{align}\label{Gegenbauer}
\mathcal{O}_{q,nk} = \partial_+^{k} \bar q\, C_{n}^{(\frac32)}  \left(\frac{\Dl_+-\Dr_+}{\Dl_++\Dr_+}\right) q\,, &&
\mathcal{O}_{g,nk} = 6\, \partial_+^{k-1} F_{+\mu} C_{n-1}^{(\frac52)}\left(\frac{\Dl_+-\Dr_+}{\Dl_++\Dr_+}\right)  F_{+\mu}\,,
\end{align}
where $C_n^{(\lambda)}(x)$ are Gegenbauer polynomials. For the $C$-parity-even operators considered here the sum goes, obviously,
over odd $n$.
 The ``coefficient functions'' $\Psi^{(\chi)}_{nk}(z_1,z_2)$ are homogeneous polynomials of two variables $z_1,z_2$ of
 degree $k$ for quarks and $k-1$ for gluons.
The factor 6 in the definition of the gluon operator is inserted for convenience.
{%
It ensures uniform normalization of the coefficient functions later on}, see Eq.~\eqref{coeff-Psi-functions}.

Renormalized LROs are defined by the same expression with bare local operators replaced by  the renormalized ones
(we always assume dimensional regularization and minimal subtraction):
\begin{align}\label{LRO-def-ren}
[\mathcal{O}_\chi(z_1,z_2)]=\sum_{n, k} \Psi^{(\chi)}_{nk}(z_1,z_2) [\mathcal{O}_{\chi,nk}(0)]\,.
\end{align}
Let us stress that the polynomials $\Psi_{nk}$ in Eq.~\eqref{LRO-def-ren} are exactly the same as in~Eq.~\eqref{LRO-def-bare}.
Note also that the LRO on the l.h.s. of Eq.~\eqref{LRO-def-ren} does not depend on the particular choice of the local operator basis.
Going over to a different basis would yield different expressions for the
polynomials {%
$\Psi_{nk}$} such that the sum is unaffected.

The renormalized local operators \eqref{Gegenbauer} satisfy the renormalization group equation (RGE) which has the usual form
\begin{align}\label{RGE-local}
\left( \Big(M\partial_M +\beta(a)\partial_a\Big)\delta^{\chi\chi'}+\gamma^{\chi\chi'}_{nn'}(a)\right)
[\mathcal{O}_{\chi',n'k}]=0 \,,
\end{align}
where $M$ is the renormalization scale
\begin{align}
  a = \frac{\alpha_s}{4\pi}\,,
\qquad
\beta(a) = M \frac{da}{dM} = -2a \big(\epsilon -\gamma_g\big) = -2a \big(\epsilon + a\beta_0 + a^2\beta_1+\ldots\big)
\end{align}
with
\begin{align}
 \beta_0 = \frac{11}{3} C_A -\frac23 n_f\,, \qquad \beta_1 = \frac23\big[17 C_A^2 -5 n_f C_A - 3 n_f C_F \big]\,.
\end{align}
The anomalous dimension matrix $\gamma^{\chi\chi'}_{nn'}$ has a triangular form in $n$-space: its elements are nonzero only for
$n\geq n'$.
The diagonal entries $\gamma_{nn}^{\chi\chi'}$ are $2\times 2$ matrices in the $\chi,\chi' \in \{q,\,g\}$ space, they are known to
three-loop accuracy~\cite{Vogt:2004mw}. The off-diagonal entries in $n$-space vanish to one-loop accuracy for the special
choice of local operators in Eq.~\eqref{Gegenbauer}. Beyond one loop the non-diagonal entries are non-zero and their calculation to
two-loop accuracy is the main topic of this study.

The RGEs for local operators of different dimension can be combined to the RGE for their generating function, the LRO.
In this representation one obtains an integro-differential equation
\begin{flalign}\label{RGE-LRO}
\Big(M\partial_M +\beta(a)\partial_a\Big)  [\mathcal{O}_\chi](z_1,z_2) & = -\mathbb{H}_{\chi\chi'} (a)[\mathcal{O}_{\chi'}](z_1,z_2),
 \end{flalign}
where $\mathbb{H}(a)$ (evolution kernel) is an integral operator which has a perturbative expansion
\begin{align}
\mathbb{H}_{\chi\chi'}(a)=  a \mathbb{H}^{(1)}_{\chi\chi'} + a^2 \mathbb{H}^{(2)}_{\chi\chi'} + \ldots\,.
\end{align}
The one-loop kernel $\mathbb{H}^{(1)}_{\chi\chi'}$ takes the form~\cite{Balitsky:1987bk}
\begin{align}
   \label{eq:Honeloop}
    \mathbb H^{(1)}
  \begin{pmatrix}
    \mathcal O_q(z_1,z_2) \\
    \mathcal O_g(z_1,z_2)
  \end{pmatrix}
  & =
  \int_\alpha
  \begin{pmatrix}
    4 C_F \bar\alpha /\alpha & 0 \\
    0 & 4C_A \bar\alpha^2/\alpha
  \end{pmatrix}
  \begin{pmatrix}
    2\mathcal O_q(z_1,z_2) - \mathcal O_q(z_{12}^\alpha,z_2)
    -\mathcal O_q(z_1,z_{21}^\alpha) \\
    2\mathcal O_g(z_1,z_2) - \mathcal O_g(z_{12}^\alpha,z_2)
    -\mathcal O_g(z_1,z_{21}^\alpha)
  \end{pmatrix}
\notag\\
  & \quad
  - \int_{\alpha\beta}
  \begin{pmatrix}
    4C_F & 4n_f z_{12} (\bar \alpha \bar \beta+3\alpha\beta)\\
    8C_Fz_{12}^{-1} & 16C_A (\bar \alpha \bar \beta+2\alpha\beta)
  \end{pmatrix}
  \begin{pmatrix}
    \mathcal O_q(z_{12}^\alpha,z_{21}^\beta) \\
    \mathcal O_g(z_{12}^\alpha,z_{21}^\beta)
  \end{pmatrix}
\notag\\
  & \quad
  + \begin{pmatrix}
    2C_F & 0 \\
    -4C_F z_{12}^{-1} & 12 C_A - 2\beta_0
  \end{pmatrix}
  \begin{pmatrix}
    \mathcal O_q(z_1,z_2) \\
    \mathcal O_g(z_1,z_2)
  \end{pmatrix}.
\end{align}
Here we introduced the shorthand notations for the integrals
\begin{align}
\int_\alpha \equiv \int_0^1 d\alpha, && \int_{\alpha\beta} \equiv \int_0^1 d\alpha d\beta\, \theta(1-\alpha-\beta).
\end{align}
The one-loop kernel commutes with the canonical generators of the (collinear) conformal transformations,
\begin{align}\label{commSH}
S_{\alpha}^{(0)} \,\mathbb{H}^{(1)}=\mathbb{H}^{(1)}\,S_{\alpha}^{(0)}\,,
 &&   (S^{(0)}_\alpha)_{\chi\chi'}=\delta_{\chi\chi'} S^{(0)}_{\chi,\alpha}.
\end{align}
where
\begin{align}
\label{S-canonical}
 S_{\chi,-}^{(0)}&=-\partial_{z_1}-\partial_{z_2}\,,
\notag\\
S_{\chi,0}^{(0)} &=z_1\partial_{z_1}+z_2\partial_{z_2}+2j_{\chi},  \qquad
\notag\\
S_{\chi,+}^{(0)} &=z_1^2\partial_{z_1}+z_2^2\partial_{z_2}+2j_{\chi}(z_1+z_2),
\end{align}
with
\begin{align}
   j_q = 1\,, \qquad j_g = 3/2\,.
\end{align}

The expressions for the one-loop kernel in Eq.~\eqref{eq:Honeloop} have been
obtained in {%
Ref.~\cite{Balitsky:1987bk} by direct calculation.
But in fact, they are determined completely by the anomalous dimensions $\gamma^{(1),\chi\chi'}_n$,
see Ref.~\cite{Bukhvostov:1985rn} }and a discussion in Ref.~\cite{Braun:2017cih}.
Starting from two loops this property is lost: The
evolution kernel does not commute any longer with the canonical generators and cannot be restored from the anomalous dimensions
alone. Nevertheless, one can simplify the calculation considerably by observing that QCD in noninteger $d-2\epsilon$ dimensions
possesses a nontrivial critical point such that the $\beta$-function vanishes for the special value of the coupling,
$\beta(a^\ast)=0$, and the theory enjoys full conformal invariance. This property allows one to argue that Eq.~\eqref{commSH} holds
true for the full kernels in arbitrary order of perturbation theory with the  appropriately modified (``deformed'') symmetry
generators $S_\alpha$.
A general technique for the calculation of quantum corrections to the generators of conformal transformations was
developed in {
~\cite{Braun:2013tva,Braun:2014vba,Braun:2016qlg}. The
present case (flavor-singlet) is more complicated as compared to the
discussion of flavor-nonsinglet operators in
{%
Ref.}~\cite{Braun:2016qlg} because of a nontrivial mixing of gauge-invariant LROs with
{%
BRST~\cite{Becchi:1975nq}}
and Equation-of-Motion (EOM) operators. We consider this problem in more detail in the next Section.

\section{Light-ray operators beyond one loop}\label{sect:LROP}

It is well known that gauge-invariant (local) operators mix under renormalization with the BRST variations and EOM operators which
are not gauge-invariant. A renormalized operator $[\mathcal{O}_{\chi,nk}]$ can be decomposed
as~\cite{Joglekar:1975nu,Joglekar:1976eb,Joglekar:1976pe,collins_1984}
\begin{align}
\label{JL}
[\mathcal{O}_{\chi,nk}(0)]= \widehat{\mathcal{ O}}_{\chi,nk} +  \mathcal{B}_{\chi,nk} + \mathcal{E}_{\chi,nk},
\end{align}
where the three terms on the r.h.s. are the gauge-invariant, BRST and EOM operators, respectively. The last two terms do not
contribute to the correlators of gauge invariant operators and in many cases can be omitted. The (renormalized) LROs can be
decomposed in a similar manner and one can expect that in most cases only the gauge-invariant contributions will prove to be
relevant,
\begin{align}
\widehat{\mathcal{O}}_\chi(z_1,z_2)=\sum_{nk} \Psi^{(\chi)}_{nk}(z_1,z_2) \widehat{\mathcal{O}}_{\chi,nk}(0)\,.
\end{align}
By definition, {the} variation of the LRO under collinear conformal transformations $\delta_\alpha$, $\alpha=0,\pm$ (or,
equivalently, dilatation and rotation in the $(n,\bar n)$ plane, $D-M_{n\bar n}$, translation along the light-cone $P_+$ and
special conformal transformation
$K_-$, cf. Appendix~\ref{App:A}) is determined by the transformation properties of local
operators~\cite{Braun:2013tva,Braun:2014vba,Braun:2016qlg}
\begin{align}\label{variation-LRO-def}
\delta_\alpha \widehat{\mathcal{O}}_\chi(z_1,z_2)   \equiv
            \sum_{nk} \Psi^{(\chi)}_{nk}(z_1,z_2) \delta_\alpha
\widehat{\mathcal{O}}_{\chi,nk}(0)\,,
\end{align}
where $\delta_\alpha\widehat{\mathcal{O}}_{\chi,nk}(0)$ can be expanded over the basis of  gauge-invariant local operators with finite
coefficients,~\footnote{
Our notation uses the following short-hands here:}
The last subscript $k+\alpha$ is meant to be $k-1$, $k$ and $k+1$ for $\alpha = -, 0, +$, respectively.}
\begin{align}\label{deltaLRO}
\delta_\alpha\widehat{\mathcal{O}}_{\chi,nk}(0)=
    \sum_{\chi',n} (c_\alpha)_{nn'}^{\chi\chi'} \widehat{\mathcal{O}}_{\chi',n',k+\alpha}(0)+\ldots\,,
\end{align}
where the ellipses stand for contributions of gauge non-invariant operators. For dilatations, the structure of such terms repeats
Eq.~\eqref{JL} -- one obtains local operators that can be presented as a BRST variation, and EOM operators. For the special
conformal transformations the structure of such gauge non-invariant contributions is more complicated, and they are in general
non-local. A detailed discussion and explicit expressions can be found in  Ref.~\cite{Braun:2018mxm} where it is proven that such
extra terms do not contribute to correlation functions with gauge-invariant operators. For the symmetry transformations from the
Poincar\'e group there are no such contributions, i.e., they involve only the sum over gauge-invariant operators.

Note that the expressions for the symmetry transformations depend on the choice of the local operators
but this dependence is compensated by the modification of the coefficient functions.
As the result, the symmetry generators in the LRO representation are defined unambiguously and do not depend
on the choice of the operator basis.

Omitting gauge non-invariant contributions and substituting Eq.~\eqref{deltaLRO} in Eq.~\eqref{variation-LRO-def}
one can represent the result as an action on the LRO of a certain linear operator
\begin{align}\label{def-Splus}
\delta_\alpha \widehat{\mathcal{O}}_\chi(z_1,z_2) =\sum_{\chi'}S_{\alpha, \chi\chi'}\, \widehat{\mathcal{O}}_{\chi'}(z_1,z_2)\,.
\end{align}
The generators $S_{\alpha, \chi\chi'}$,  $\alpha = \pm, 0$  are integro-differential operators in $z_1,z_2$, which are $2\times 2$ matrices in the
quark-gluon space, {%
$\chi,\chi' \in \{q,\,g\}$.}
From general considerations~\cite{Braun:2013tva,Braun:2016qlg} it follows that they can be written in the form
\begin{align}\label{Salpha}
S_{-} & = S_-^{(0)}\,,
\notag\\
S_{0} \,& = S^{(0)}_{0}   -\epsilon\,
+ \frac12 \mathbb{H}(a_*)\,,
\notag\\
S_{+} & = S^{(0)}_{+} +
(z_1+z_2)\Big(-\epsilon
+ \frac12 \mathbb{H}(a_*)\Big)
    +(z_1-z_2) \Delta(a_*)\,,
\end{align}
where the canonical generators, $S^{(0)}_\alpha$, are defined  in {%
Eq.}~\eqref{S-canonical}. Thus quantum corrections to the classical
symmetry generators  involve the evolution kernel
$\mathbb{H}$ and the operator $\Delta$ which we will refer to as conformal anomaly.
Both of them are matrices in the quark-gluon space.

\subsection{Flavor-singlet conformal anomaly $\Delta$}

The leading-order (one-loop) conformal anomaly  for the case under consideration
\begin{align}
  \Delta(a_\ast) = a_\ast \Delta^{(1)} + \mathcal O (a_\ast^2)
\end{align}
has been calculated in Ref.~\cite{Belitsky:1998gc} from the analysis of the scale and conformal Ward identities.
Our approach is in principle similar but differs from {
~\cite{Belitsky:1998gc} in details. The calculation
is described in Appendix~\ref{App:B}. We obtain
\begin{align}\label{Delta-1-loop}
  \Delta^{(1)}
    \begin{pmatrix}
    \mathcal O_q(z_1,z_2) \\
    \mathcal O_g(z_1,z_2)
  \end{pmatrix}
&= -2\int_\alpha
  \begin{pmatrix}
     C_F\! \left[\frac{\bar \alpha}{\alpha} + \ln \alpha\right] & 0 \\
    z_{12}^{-1} C_F  & C_A \left[\frac{\bar \alpha}{\alpha} +2 \ln \alpha\right]
  \end{pmatrix}
  \begin{pmatrix}
    \mathcal O_q(z_{12}^\alpha,z_2) -\mathcal O_q(z_1,z_{21}^\alpha) \\
    \mathcal O_g(z_{12}^\alpha,z_2) -\mathcal O_g(z_1,z_{21}^\alpha)
  \end{pmatrix}\,. &
\end{align}
This expression is in agreement with Ref.~\cite{Belitsky:1998gc}.

The general procedure how one can restore the evolution kernels from the
spectrum of anomalous dimensions at the same order in perturbation theory
and the conformal anomaly at one order less is explained in detail
in {
~\cite{Braun:2017cih} and it can be used for the flavor-singlet operators
as well.

The central  observation is that the evolution kernel at critical coupling
must commute with the generators of conformal transformations,
\begin{align}
  [\mathbb H(a_\ast) , S_+(a_\ast) ] = 0\,,
\end{align}
where both operators are now $2\times2$-matrices. Expanding this equation in powers of the  coupling yields a set of nested
equations, e.g. up to $\mathcal O(a_\ast^2)$,
\begin{subequations}\label{eq:Hconstraint}
\begin{align}
  [ S_+^{(0)}, \mathbb H^{(1)} ] &= 0\,,
  \label{eq:Hconstraint1loop} \\
  [ S_+^{(0)}, \mathbb H^{(2)} ] &=  [ \mathbb H^{(1)}, z_{12}\Delta^{(1)} ] +
  [ \mathbb H^{(1)},(z_1+z_2) ]\left(-\epsilon + \frac12 \mathbb H^{(1)}\right)\,.
  \label{eq:Hconstraint2loop}
\end{align}
\end{subequations}
Eq.~\eqref{eq:Hconstraint1loop} is a first-order homogeneous differential equation
and its general solution (which we call ``invariant operator'' in what follows)
takes the generic form
\begin{align}
  \label{eq:Hinvgeneric}
  \mathbb H_{\rm inv}
  \begin{pmatrix}
    \mathcal O_q(z_1,z_2) \\
    \mathcal O_g(z_1,z_2)
  \end{pmatrix}
  =
  \int_{\alpha\beta}
  \begin{pmatrix}
    h_{\rm inv,qq}(\tau) & z_{12} \bar \alpha \bar \beta h_{\rm inv,qg}(\tau)\\
    z_{12}^{-1}h_{\rm inv,gq}(\tau) & \bar \alpha \bar \beta h_{\rm inv,gg}(\tau)
  \end{pmatrix}
  \begin{pmatrix}
    \mathcal O_q(z_{12}^\alpha,z_{21}^\beta) \\
    \mathcal O_g(z_{12}^\alpha,z_{21}^\beta)
  \end{pmatrix}\,,
\end{align}
where $h_{\rm inv}(\tau)$ are functions of $\tau = \frac{\alpha\beta}{\bar \alpha \bar\beta}$ dubbed ``the conformal ratio''.
Indeed, the one-loop evolution kernel Eq.~\eqref{eq:Honeloop} can be rewritten in this form~\cite{Braun:2014vba}.
Eq.~\eqref{eq:Hconstraint2loop}, in turn, can be viewed as an \emph{inhomogeneous} differential equation for  the two-loop
evolution kernel $\mathbb H^{(2)}$ and is solved by~\cite{Braun:2017cih}
\begin{align}\label{eq:H2loop}
  \mathbb H^{(2)} = \mathbb H_{\rm inv}^{(2)} +
  [ \mathbb H^{(1)}, \mathbb X^{(1)} ] +
  \mathbb T^{(1)}\left(\gamma_g + \frac12 \mathbb H^{(1)}\right)\,,
\end{align}
where $\mathbb H_{\rm inv}^{(2)}$ can be any invariant operator that takes the form Eq.~\eqref{eq:Hinvgeneric} and the
operators $\mathbb X^{(1)}$ and $\mathbb T^{(1)}$ are $2\times2$-matrices which must satisfy the constraints
\begin{align}\label{eq:X1T1constraint}
    [S_+^{(0)} , \mathbb X^{(1)}] = z_{12} \Delta^{(1)}\,, &&
    [S_+^{(0)} , \mathbb T^{(1)} ] = [\mathbb H^{(1)} , z_1+z_2]\,.
\end{align}
Solving these two equations we get
\begin{align}
\label{eq:Xoneloop}
 \mathbb X^{(1)}_{qq} \mathcal O_q (z_1,z_2) &=
2 C_F \int_\alpha \frac{\ln\alpha}{\alpha}
\Big(2\mathcal O_q(z_1,z_2)- \mathcal O_q(z_{12}^\alpha,z_2) -\mathcal O_q(z_1,z_{21}^\alpha)\Big)\,,
\notag\\
 \mathbb X^{(1)}_{gg} \mathcal O_g (z_1,z_2) &=
2 C_A \int_\alpha \frac{\ln\alpha}{\alpha}
\Big(2\mathcal O_g(z_1,z_2)- \mathcal O_g(z_{12}^\alpha,z_2) -\mathcal O_g(z_1,z_{21}^\alpha)\Big)\,,
\notag\\
\mathbb X^{(1)}_{gq} \mathcal O_q (z_1,z_2) &=
 -2C_F \frac1{z_{12}}\int_\alpha\Big(\mathcal O_q(z_{12}^\alpha,z_2) +\mathcal O_q(z_1,z_{21}^\alpha)\Big)\,,
\notag\\
\mathbb X^{(1)}_{qg} \mathcal O_g (z_1,z_2) &=0\,.
\end{align}
The operator $\mathbb T^{(1)}$ takes the form
\begin{align}
  \label{eq:Toneloop}
  \mathbb T^{(1)}
  \begin{pmatrix}
    \mathcal O_q(z_1,z_2) \\
    \mathcal O_g(z_1,z_2)
  \end{pmatrix}
  &=
  -\int_{\alpha\beta} \ln(1-\alpha-\beta)
  \begin{pmatrix}
    4C_F \, & 4n_f z_{12} (\bar \alpha \bar \beta+3\alpha\beta)\\
    8C_Fz_{12}^{-1}\, & 16C_A (\bar \alpha \bar \beta+2\alpha\beta)
  \end{pmatrix}
  \begin{pmatrix}
    \mathcal O_q(z_{12}^\alpha,z_{21}^\beta) \\
    \mathcal O_g(z_{12}^\alpha,z_{21}^\beta)
  \end{pmatrix}
  \notag \\
  & \quad-
  \int_\alpha \ln(\bar \alpha)
  \begin{pmatrix}
    4 C_F \bar\alpha /\alpha & 0 \\
    0 & 4C_A \bar\alpha^2/\alpha
  \end{pmatrix}
  \begin{pmatrix}
    \mathcal O_q(z_{12}^\alpha,z_2) +\mathcal O_q(z_1,z_{21}^\alpha) \\
    \mathcal O_g(z_{12}^\alpha,z_2) +\mathcal O_g(z_1,z_{21}^\alpha)
  \end{pmatrix}\,. &
\end{align}
This expression differs from the evolution kernel in Eq.~\eqref{eq:Honeloop} only by the insertion of a factor
$\ln(1-\alpha-\beta)$. Note that the constraints in Eq.~\eqref{eq:X1T1constraint} fix  $\mathbb X^{(1)}$ and  $\mathbb T^{(1)}$
only up to contributions of invariant operators. Such possible extra terms enter the two-loop kernel $\mathbb H^{(2)}$ through the
product with $\mathbb H^{(1)}$ which is itself an invariant operator. One can easily verify that the product of two invariant
operators Eq.~\eqref{eq:Hinvgeneric} is again an invariant operator.  Therefore, the freedom in the definition of  $\mathbb
X^{(1)}$ and $\mathbb T^{(1)}$ corresponds to a redefinition of the two-loop invariant evolution kernel $\mathbb H_{\rm inv}^{(2)}$
which still has to be determined.

The expressions given above define the non-invariant part of the two-loop evolution kernels
in the factorized form
\begin{align}
\Delta \mathbb{H}^{(2)}=[ \mathbb H^{(1)}, \mathbb X^{(1)} ] +
  \mathbb T^{(1)}\left(\gamma_g + \frac12 \mathbb H^{(1)}\right)
\end{align}
as a product of relatively simple integral operators~\eqref{eq:Xoneloop},~\eqref{eq:Toneloop}.
For certain applications and in particular for numerical studies of the scale dependence of the GPDs,
it can be advantageous to have explicit expressions for these products.

The results can be written in the form
\begin{align}
\Delta\mathbb{H}=\begin{pmatrix}
\phantom{{z_{12}^{-1}}}\Delta\mathbb{H}_{qq} & z_{12}\Delta\mathbb{H}_{qg}
\\
{z_{12}^{-1}}\,\Delta\mathbb{H}_{gq} & \phantom{z_{12}}\Delta\mathbb{H}_{gg}
\end{pmatrix}\,,
\end{align}
where
\begin{align}
\Delta \mathbb{H} f(z_1,z_2)=\Delta r f(z_1,z_2)+\int_\alpha \vartheta(\alpha)[ 2 f(z_1,z_2)-f(z_{12}^\alpha,z_2)-f(z_1,z_{21}^\alpha)]
+ \int_{\alpha\beta} \Delta\omega(\alpha,\beta) f(z_{12}^\alpha,z_{21}^\beta)\,.
\end{align}
For the pure singlet-contribution to the quark kernel we get $\Delta r_{qq}^{\rm PS}=\vartheta_{qq}^{\rm PS}=0$ and
\begin{align}\label{NI-PS}
\Delta\omega_{qq}^{\rm PS}(\alpha,\beta) & =8C_F n_f\Biggl\{\frac{14\tau}{\bar\tau}+\ln^2(1-\alpha-\beta)-\ln^2\bar\alpha-\ln^2\bar\beta
+\left(1+\frac{8\tau}{\bar\tau}\right)\ln(1-\alpha-\beta)
\notag\\
&\quad
-(1-8\alpha)\ln\left(1-\frac\beta{\bar\alpha}\right) -(1-8\beta)\ln\left(1-\frac\alpha{\bar\beta}\right)
\Biggr\}\,,
\end{align}
and for the non-singlet contribution (cf.~Ref.~\cite{Braun:2014vba})
\begin{align}
\Delta r_{qq}& = 8C_F\left(\beta_0\left(\frac{\pi^2}6-1\right)+C_F\left(3-\frac{\pi^2}6\right)\right)
=-2\int_0^1 d\alpha \, \vartheta_{qq} (\alpha)\,,
\notag\\
\vartheta_{qq} (\alpha)& = 8C_F\frac{\bar\alpha}{\alpha}\ln\bar\alpha\left(\frac12\beta_0-C_F\left(\frac32-\ln\bar\alpha
+\frac{1+\bar\alpha}{\bar \alpha}\ln\alpha\right)\right)\,,
\notag\\
\Delta\omega_{qq}(\alpha,\beta) &=4C_F\Bigg\{-\beta_0\ln(1-\alpha-\beta)
+C_F\bigg(-\ln^2\bar\alpha-\ln^2\bar\beta-\ln^2(1-\alpha-\beta)
\notag\\
&\quad +2\ln\alpha(\ln\bar\alpha+1)+2\ln\beta(\ln\bar\beta+1)
-\frac2\alpha{\ln\bar\alpha}-\frac2\beta{\ln\bar\beta}+\ln(1-\alpha-\beta)\bigg)
\Bigg\}\,.
\end{align}
For the off-diagonal kernels we obtain $\Delta r_{qg}=\vartheta_{qg}=0$,
\allowdisplaybreaks{
\begin{flalign}
\Delta\omega_{qg}(\alpha,\beta)&=
    8n_f\bar\alpha\bar\beta
        \Biggl\{    C_F \Biggl(  (1+5\tau)\ln\tau-9\tau +
    \frac12\bar\tau\Big(\ln^2(1-\alpha-\beta)-\ln^2\bar\alpha-\ln^2\bar\beta\Big) &
\notag\\
&\quad
    -\frac12(1+3\tau)\left(\ln^2\frac\alpha{\bar\alpha}+\ln^2\frac{\beta}{\bar\beta}\right)
    -(1+\tau)\left(\frac1{\bar\alpha}\ln\alpha+\frac1{\bar\beta}\ln\beta\right)
\Biggr)
\notag \\
&\quad
    -C_A\Biggl(     \frac{25}3 \tau +\frac{4}{3}(\bar\tau-\tau) \ln\bar\tau +3\tau\ln\tau
+
    (3+\tau)\Big(\ln ^2(1-\alpha-\beta)-\ln ^2\bar\alpha-\ln ^2\bar\beta\Big)
\notag\\
&\quad
    -\frac12(1+3\tau)\Big(\ln^2\alpha+\ln^2\beta-\ln^2\bar\alpha-\ln^2\bar\beta\Big)
+\frac{8\alpha}{\bar\beta}\ln\left(1-\frac\beta{\bar\alpha}\right)
+\frac{8\beta}{\bar\alpha}\ln\left(1-\frac\alpha{\bar\beta}\right)
\notag\\
&\quad
+\frac{2\beta}{\bar\alpha\bar\beta}\ln\bar\alpha + \frac{2\alpha}{\bar\beta\bar\alpha}\ln\bar\beta
    -\frac{\beta}{\bar\beta}\ln\frac{\beta}{\bar\alpha}-\frac{\alpha}{\bar\alpha}\ln\frac{\alpha}{\bar\beta}
    -\bar\tau\ln(1-\alpha-\beta)
\Biggl)
\Biggr\}\,,
\end{flalign}
} and
\begin{flalign}
\Delta r_{gq} & = 8 C_F\Big( \beta_0 -2 C_A
-C_F
\Big), &
\notag\\
\vartheta_{gq}(\alpha) & = 8 C_F \Biggl\{
-\frac12\beta_0 + C_A\left(\frac1\alpha\ln\frac\alpha{\bar\alpha}+\frac\alpha{\bar\alpha}\ln\alpha\right)
+C_F
\left(\frac32-\frac{\bar\alpha}\alpha\ln\alpha\right)
\Biggr\}\,,
\notag\\
\Delta \omega_{gq} &=  8 C_F\Biggl\{
-\beta_0\ln(1-\alpha-\beta)
+ C_A \Biggl(\left[\frac1\alpha\right]_+ + \left[\frac1\beta\right]_+ + 42\frac\tau {\bar\tau}
+ \frac\alpha{\bar\alpha}+\frac\beta{\bar\beta}
   -\ln^2\frac\alpha{\bar\alpha}-\ln^2\frac\beta{\bar\beta}
\notag\\
&\quad
+2\Big(\ln^2(1-\alpha-\beta)-\ln^2\bar\alpha-\ln^2\bar\beta\Big)
        -\frac\alpha{\bar\alpha}\left(4+\frac\alpha{\bar\alpha}\right)\ln\alpha
        -\frac\beta{\bar\beta}\left(4+\frac\beta{\bar\beta}\right)\ln\beta
\notag\\
&\quad
-3(1-8\alpha)\ln\left(1-\frac\beta{\bar\alpha}\right)-3(1-8\beta)\ln\left(1-\frac\alpha{\bar\beta}\right)
+24\frac{\tau}{\bar\tau}\ln(1-\alpha-\beta)
\Biggr)
\notag\\
&\quad
+C_F\left(
2-\left[\frac1\alpha\right]_+-\left[\frac1\beta\right]_+
    +\ln^2\alpha+\ln^2\beta-\ln^2(1-\alpha-\beta)+3\ln(1-\alpha-\beta) -2\ln\bar\tau
\right)\Biggr\}\,,
\end{flalign}
where the ``plus'' distribution is defined as
\begin{align}
\int_{\alpha\beta} \left[\frac1\alpha\right]_+ f(z_{12}^\alpha,z_{21}^\beta)\equiv
\int_{\alpha\beta} \frac1\alpha \Big(f(z_{12}^\alpha,z_{21}^\beta)- f(z_{1},z_{21}^\beta)\Big)\,.
\end{align}
Finally, for the gluon-gluon kernel we obtain
\begin{align}\label{NI-GG}
\Delta r_{gg} & = 4 C_A^2 \Big(-3+\pi^2\Big)=-2\int_0^1d\alpha\, \vartheta_{gg}(\alpha)\,,
\notag\\
\vartheta_{gg}(\alpha) &= 8C_A^2\,\frac{\bar\alpha^2}\alpha{\ln\bar\alpha} \left( \ln\bar\alpha
-\frac{1+\bar\alpha^2}{\bar\alpha^2}\ln\alpha\right)\,,
\notag\\
 \Delta w_{gg}(\alpha,\beta) & = 8 C_F n_f\Biggl\{
(1-\alpha-\beta)\Big(\ln^2(1-\alpha-\beta)-\ln^2\bar\alpha-\ln^2\bar\beta\Big)+4\alpha\ln\bar\beta+4\beta\ln \bar\alpha
\notag\\
&\quad +6 \alpha\beta  -  \left(\frac\alpha{\bar\alpha} +\frac\beta{\bar\beta}\right) (\bar\alpha\bar\beta+\alpha\beta)
    \Biggr\}
    + 16 C_A^2\Biggl\{ \bar\alpha\bar\beta\Biggl[
   - 32\tau +2\tau\ln\bar\tau+(1+4\tau)\ln\tau
\notag\\
&\quad
    +2(1+2\tau)\left(\ln\bar\alpha\ln\frac\alpha{\bar\alpha}+\ln\bar\beta\ln\frac\beta{\bar\beta}\right)
     -2 (2+\tau)\Big(\ln^2(1-\alpha-\beta)-\ln^2\bar\alpha-\ln^2\bar\beta\Big)
\notag\\
&\quad
-12\left(\frac{\alpha}{\bar\beta}\ln\left(1-\frac\beta{\bar\alpha}\right)
+\frac{\beta}{\bar\alpha}\ln\left(1-\frac\alpha{\bar\beta}\right)\right)
-\frac12\tau\left(\frac{1+4\bar\alpha}{\bar\alpha^2}\ln\alpha+\frac{1+4\bar\beta}{\bar\beta^2}\ln\beta\right)
  \Biggr]
  \notag\\
&\quad
-\left(6\beta-\frac{\beta}{2\alpha}+\frac{\bar\alpha}{\alpha}\right)\ln\bar\alpha
-\left(6\alpha-\frac{\alpha}{2\beta}+\frac{\bar\beta}{\beta}\right)\ln\bar\beta
+\frac{\alpha}{\bar\beta}+\frac{\beta}{\bar\alpha}
\Biggr\}\,.
\end{align}
The expressions in Eqs.~\eqref{NI-PS}-\eqref{NI-GG} supplemented by the invariant kernels which are calculated in the next Section
provide one with the complete two-loop flavor-singlet evolution kernels in the LRO representation. These are our main results.

\section{Anomalous dimensions vs invariant kernels}\label{sec:inv-kernels}

Substituting the LRO in Eq.~\eqref{RGE-LRO} by its expansion in  terms of local operators and comparing the resulting expression
with the RGE~\eqref{RGE-local} one gets
\begin{align}
[\mathbb{H} \Psi]^{\chi}_{nk}(z_1,z_2)=\sum_{n'\geq n}\Psi^{\chi'}_{n', k}(z_1,z_2) \gamma_{n'n}^{\chi'\chi}\,.
\end{align}
Conformal operators of the lowest dimension for the given spin correspond to {%
the highest weights}
of the representation and are annihilated by $S_-= S_-^{(0)}$.
As a consequence, the coefficient functions of the operators with $k=n$ are translation-invariant
and by dimension counting
$\Psi_{nn}^{q}\sim z_{12}^{n}$ and $\Psi^{g}_{nn}\sim z_{12}^{n-1}$.
It follows that
\begin{align}
 \mathbb{H}_{\chi\chi'} e^{\chi'}_n=h_{\chi\chi'}(n) e^\chi_n\,, \qquad
  e^q_n=\begin{pmatrix}z_{12}^{n} \\ 0 \end{pmatrix}\,,\qquad
e^g_n=\begin{pmatrix}0\\z_{12}^{n-1} \end{pmatrix}\,,
\end{align}
where the coefficients $h_{\chi\chi'}$ are related to the matrix of anomalous dimensions $\gamma_{\chi\chi'}(n)$ in the
normalization {%
chosen in Ref.~\cite{Vogt:2004mw} as}
\begin{align}\label{eq:h2gamma}
h_{\chi\chi}(n)=2 \gamma_{\chi\chi}^{\text{{\color{blue}
Ref.}{\cite{Vogt:2004mw}}}}(n+1)
\,, &&
h_{q g}(n)= (2/ n) \gamma_{q g}^{\text{{\color{blue}Ref.}{\cite{Vogt:2004mw}}}}(n+1)
\,, &&
h_{gq}(n)= 2 n\, \gamma_{gq}^{\text{{\color{blue}Ref.}{\cite{Vogt:2004mw}}}}(n+1)\,.
\end{align}
Eq.~\eqref{eq:Hconstraint2loop} allows one to find the spectrum of the invariant kernel $\mathbb H_{inv}^{(2)}$ from the known
results for the two-loop anomalous dimensions. Following the notation in Ref.~\cite{Vogt:2004mw} we split the two-loop
flavor-singlet quark kernel in two parts -- flavor non-singlet and pure-singlet, $\mathbb H^{(2)}_{qq}=\mathbb H^{(2)}_{\rm
NS}+\mathbb H^{(2)}_{\rm PS}$. The expression for the non-singlet part can be found in {Ref.}~\cite{Braun:2017cih}, while for the
pure singlet one finds
\begin{align}\label{eq:pstwoloopspectrum}
h^{(2)}_{\rm PS}(n)=h^{(2),\text{inv}}_{\rm PS}(n) + h^{(1)}_{qg}(n) X^{(1)}_{gq}(n)+\frac12 h^{(1)}_{gq}(n) \partial_n h^{(1)}_{qg}(n)\,.
\end{align}
where $h^{(2)}_{\rm PS}(n)$ and $ h^{(1)}_{\chi\chi'}(n)$ are related to the corresponding anomalous dimensions as in
Eq.~\eqref{eq:h2gamma}, $\partial_n = \frac{\partial}{\partial n}$ and
\begin{align}\label{eq:X1gq}
 \mathbb{X}_{\chi\chi'} e^{\chi'}_n=X_{\chi\chi'}(n) e^\chi_n\,,
 \qquad
{%
X^{(1)}_{gq}(n)}=-\frac{4C_F}{(n+1)}\,.
\end{align}
Using
\begin{align}
\label{h1qg+gq}
h^{(1)}_{qg}(n)=-4 n_f\frac{(n+2)(n+1)+2}{n(n+1)(n+2)(n+3)}\,, &&
h^{(1)}_{gq}(n)=-4 C_F\frac{(n+2)(n+1)+2}{(n+1)(n+2)}\,,
\end{align}
and the two-loop expression for the pure-singlet anomalous dimension from Ref.~\cite{Vogt:2004mw}, we get
\begin{align}\label{PS-spectrum}
h^{(2),inv}_{\rm PS}(n)=16C_F n_f\left(-\frac{11}{n(n+1)(n+2)(n+3)}+ \frac2{(n+2)^2(n+1)^2}+\frac{6}{n^2(n+3)^2}\right)\,.
\end{align}
The corresponding kernel as a function of the conformal ratio
\begin{align}
\mathbb H_{\rm PS}^{\text{inv}} \mathcal{O}_q(z_1,z_2)=
            \int_{\alpha\beta} \omega_{\rm PS}(\tau)\mathcal{O}_q(z_{12}^\alpha,z_{21}^\beta)
\end{align}
takes the form
\begin{align}\label{pure-singlet}
\omega_{\rm PS}(\tau)=-64C_F n_f\frac1{\bar\tau}\left(\ln\bar \tau  +\frac74\tau \right)\,.
\end{align}

In a similar manner one obtains from Eq.~\eqref{eq:Hconstraint2loop} the
equations for the remaining entries:
\begin{align}\label{eq:twoloopspectrum}
h_{qg}^{(2)}(n) &=h^{\text{inv}}_{qg}(n) +h_{qg}^{(1)}(n) \left[X_{gg}^{(1)}(n)-X^{(1)}_{qq}(n)
                  + \frac12 \partial_n h_{qq}^{(1)}(n)\right]
+\left[\beta_0+\frac12 h_{gg}^{(1)}(n)\right] \partial_n h_{qg}^{(1)}(n)\,,
\notag\\
h_{gq}^{(2)}(n) &=h^{\text{inv}}_{gq}(n) +h_{gq}^{(1)}(n) \left[X_{qq}^{(1)}(n)-X^{(1)}_{gg}(n)
            + \frac12 \partial_n h_{gg}^{(1)}(n)\right]
+\left[\beta_0+\frac12 h_{qq}^{(1)}(n)\right] \partial_n h_{gq}^{(1)}(n)
\notag\\
&\quad +X_{gq}^{(1)}(n)\Big(h_{gg}^{(1)}(n)-h_{qq}^{(1)}(n)\Big)\,,
\notag\\
h_{gg}^{(2)}(n) &=h^{\text{inv}}_{qg}(n) +\left[-X^{(1)}_{gq}(n)+\frac12 \partial_n h_{gq}^{(1)}(n)\right]h_{qg}^{(1)}(n)
+
\partial_n h_{gg}^{(1)}(n)  \left(\beta_0+\frac12h_{gg}^{(1)}(n)\right)\,.
\end{align}
The relevant eigenvalues of the $\mathbb{X}$ kernels are equal to
{%
\begin{align}\label{eq:X-spectrum}
  X_{qq}^{(1)}(n)=-4C_F S_{1,1}(n)\,,
  &&
  X_{gg}^{(1)}(n)=-4 C_A S_{1,1}(n-1)\,,
\end{align}
cf. Eq.~\eqref{eq:X1gq} for $X^{(1)}_{gq}$,}
and the eigenvalues for the diagonal entries of the one-loop evolution kernels take the familiar form
\begin{align}
\label{h1qq+gg}
h_{qq}^{(1)}(n)&=4C_F\left(S_1(n+2)+S_1(n)-\frac32\right)\,,
\notag\\
h_{gg}^{(1)}(n)&=8C_A\left(S_1(n+1)-\frac{2((n+2)(n+1)+1)}{n(n+1)(n+2)(n+3)}\right)-2\beta_0\,.
\end{align}
Their derivatives can be written as
\begin{align}
\label{h1derivative}
\partial_n h_{qq}^{(1)}(n)&=4C_F\left(-S_2(n+2)-S_2(n)+\frac{\pi^2}3\right)\,,
\notag\\
\partial_n h_{gg}^{(1)}(n)&=8C_A\left(-S_2(n+1)+\frac{\pi^2}6-\partial_n\frac{2((n+2)(n+1)+1)}{n(n+1)(n+2)(n+3)}\right)\,.
\end{align}
Combining all these expressions we obtain, after some algebra, the following results for the eigenvalues of the invariant kernels
\begin{align}\label{GQ-spectrum}
h_{gq}^{\text{inv}}(n) &=%
8 C_F \Biggl\{ \beta_0\Biggl[   \left(1+\frac2{(n+2)(n+1)}\right) S_1(n+1)
-\frac83 -\frac{25}3\frac1{(n+1)(n+2)} -\frac1{(n+1)^2(n+2)^2}
\Biggr]
\notag\\
&\quad
-C_F\Biggl[   \left(5+\frac{16}{(n+2)(n+1)}\right) S_1(n+1)
-6 -\frac{25}2\frac1{(n+1)(n+2)} -\frac3{(n+1)^2(n+2)^2}\Biggr]
\notag\\
&\quad + 2 C_A\Biggl[
\left(1+\frac2{(n+2)(n+1)}\right) \left(S_{-2}(n+1)+\frac{\pi^2}6\right)+\left(1+\frac5{(n+2)(n+1)}\right) S_{1}(n+1)
\notag\\
&\quad-\frac76 +\frac{20}3\frac1{(n+1)(n+2)}
-\frac{13}2\frac1{n(n+3)}+\frac1{(n+1)^2(n+2)^2} +\frac{18}{n^2(n+3)^2}\Biggr]
\Biggr\}\,,
\end{align}
\begin{align}\label{QG-spectrum}
h_{qg}^{\text{inv}}(n) &= -4C_F\frac{\pi^2}6 h_{qg}^{(1)}(n)
+32C_F n_f\Biggl\{\frac{S_1(n+1)}{n(n+1)(n+2)(n+3)}+\frac54\frac{1}{(n+1)(n+2)}
\notag\\
&\quad -\frac{15}8\frac1{n(n+3)}
-\frac18\frac1{(n+1)^2(n+2)^2}\Biggr\}
\notag\\
&\quad
+16 C_A n_f\Biggl\{\left(1+\frac{4}{n(n+3)}\right)\frac{S_{-2}(n+1)}{(n+1)(n+2)}
      -\frac{2 S_1(n+1)}{n(n+1)(n+2)(n+3)}+\frac{18}{n^3(n+3)^3}
\notag
\\
&\quad
-\frac{29}{4}\frac{1}{(n+1)(n+2)}
+\frac{27}4\frac1{n(n+3)}-\frac2{(n+1)^2(n+2)^2} - \frac{15}{n^2(n+3)^2}
\Biggr\}\,,
%
\end{align}
and
\begin{align}\label{GG-spectrum}
h_{gg}^{\text{inv}}(n) &=\frac{16}{3}C_A n_f\Biggl\{
-\frac53 S_1(n+1)+1+ \frac1{(n+1)(n+2)}\left[\frac{19}3-\frac1{(n+1)(n+2)}+\frac{23}{n(n+3)}\right]
\Biggr\}
\notag \\
&\quad + 8C_F n_f \Biggl\{
\frac12 +\frac{2}{(n+1)(n+2)} -\frac{5}{(n+1)^2(n+2)^2} +\frac{10}{n(n+1)(n+2)(n+3)}
\Biggl\}
\notag\\
&\quad +16C_A^2\Biggl\{-\frac43 +\frac{67}{18} S_1(n+1)+S_{-3}(n+1)-2S_{1,-2}(n+1)-S_3(n+1)
\notag\\
&\quad +\frac{4((n(n+3)+3)}{n(n+1)(n+2)(n+3)} S_{-2}(n+1) + \frac{23}{36}\frac1{(n+1)(n+2)}-\frac{40}3\frac{1}{n(n+3)}
\notag\\
&\quad
+\frac56\frac1{(n+1)^2(n+2)^2}-\frac{15}{n^2(n+3)^2}
+\frac{54}{n^3(n+3)^3}
\Biggr\}\,.
\end{align}

It remains to restore the invariant kernels in the LRO representation from these results for the spectrum. The general expression
for $\mathbb{H}^{\text{inv}}$ can be written as a $2\times 2$ matrix
\begin{align}
 \mathbb{H}^{\text{inv}}=\begin{pmatrix}
\phantom{{z_{12}^{-1}}}\mathbb{H}^{\text{inv}}_{qq} & z_{12}\mathbb{H}^{\text{inv}}_{qg}
\\
{z_{12}^{-1}}\,\mathbb{H}^{\text{inv}}_{gq} & \phantom{z_{12}}\mathbb{H}^{\text{inv}}_{gg}
\end{pmatrix}\,.
\end{align}
The entries are invariant operators  which we parameterize as follows
\begin{align}\label{three-terms-expansion}
\mathbb{H}^{\text{inv}}_{\chi q}f(z_1,z_2) &= \Gamma_{\chi q} \int_\alpha \frac{\bar\alpha}{\alpha}
    \Big[2f(z_1,z_2)-f(z_{12}^\alpha,z_2)-f(z_1,z_{21}^\alpha)\Big]
+ r_{\chi q} f(z_1,z_2)\! +\!\int_{\alpha\beta} \omega_{\chi q}(\tau) f(z_{12}^\alpha,z_{21}^\beta)\,,
\notag\\
\mathbb{H}^{\text{inv}}_{\chi g}f(z_1,z_2) &= \Gamma_{\chi g} \int_\alpha \frac{\bar\alpha^2}{\alpha}
    \Big[2f(z_1,z_2)-f(z_{12}^\alpha,z_2)-f(z_1,z_{21}^\alpha)\Big]
+ r_{\chi g} f(z_1,z_2)
\notag\\
&\quad +\int_{\alpha\beta} \bar\alpha\bar\beta \,\omega_{\chi g}(\tau) \,f(z_{12}^\alpha,z_{21}^\beta)\,,
\end{align}
where $\tau =\alpha\beta/\bar\alpha \bar\beta$. Adding the pure-singlet kernel~\eqref{pure-singlet} to the expression for the flavor
non-singlet kernel given in {%
Ref.}~\cite{Braun:2014vba,Braun:2017cih} one obtains
\begin{subequations}
\begin{align}\label{chi-inv-2}
 \Gamma_{qq} &= 16 C_F \left[C_A  \left(\frac{67}{36}-\frac{\pi^2}{12}\right)- \frac5{18} n_f \right]
=\frac43 C_F \Big[C_A (4-\pi^2)+ 5\beta_0 \Big],
\\
 r_{qq} & = \frac13C_F\biggl[ \beta_0 \big(37-4\pi^2\big) +
                         C_F\big (43-4\pi^2\big ) +\frac1{N_c}\big (26-8\pi^2+72\zeta_3\big ) \biggr]\,,
\\
\omega_{qq}(\tau)& = 4C_F\biggl [ -\frac{11}3 \beta_0
+ C_F\left(\ln\bar\tau - \frac{20}3+\frac{2\pi^2}3\right)
-\frac2{N_c}\left(\Li_2(\tau)
+\left(\tau-\frac1\tau\right)\ln\bar\tau-\frac{\pi^2}6+\frac53\right)
\notag\\
&\quad -16 n_f\frac1{\bar\tau}\left(\ln\bar \tau  +\frac74\tau \right)
\biggr ]\,,
\end{align}
\end{subequations}
{%
where $\frac1{N_c}=C_A-2C_F$ for an SU$(N_c)$ gauge theory.}
For the gluon-quark kernel one obtains using the expressions collected in Appendix~\ref{App:D},
\begin{subequations}
\begin{align}
\Gamma_{gq} &= 4C_F\Big[2C_A - 5 C_F+ \beta_0
\Big]\,,
\\
r_{gq} &=8C_F\Big[C_F + \left(\frac{\pi^2}6-\frac13\right)C_A  -\frac53\beta_0\Big]\,,
\\
\omega_{gq}(\tau) &=8C_F\Biggl[ \beta_0\left(-\frac{25}3+\ln\bar\tau-\ln\tau\right)
+ C_F\left(\frac{25}2 - 3\ln\bar\tau + 8\ln\tau\right)
\notag\\
&\quad +2C_A\left(\frac{\pi^2}6+\frac16 - \frac{\tau}{\bar\tau} +\frac12 \bar\tau +5\ln\bar\tau-\frac52\ln\tau
    -12 \frac{\ln\bar\tau}{\bar\tau} +\Li_2(\tau)\right)
\Biggr]\,.
\end{align}
\end{subequations}
For the quark-gluon kernel we find  $\Gamma_{qg}=r_{qg}=0$ and
\begin{align}
\omega_{qg}(\tau) & = 16 n_f \Biggl\{
 C_F \left[\frac{\pi^2}{6}(1+3\tau)-\frac54-\frac{17}4 \tau -\tau\ln\tau +\frac14\bar\tau\ln\bar\tau \right]
    + C_A \Biggl[
-\frac{\pi^2}{12}(1+3\tau)
\notag\\
&\quad
+\frac12(1+3\tau)\Li_2(\tau)+(1+\tau)\ln\bar\tau\left(\ln\bar\tau+\frac53\right)
+ \tau\ln\tau+4\ln\bar\tau + \frac23\tau-\frac12 \Biggr]
 \Biggr\}\,.
\end{align}
Finally, for the gluon-gluon kernel we obtain
\begin{align}
\Gamma_{gg}&=-\frac{40}9 C_A n_f +\frac49\big( 67-3\pi^2\big) C_A^2=\frac43 C_A\Big[(4-\pi^2)C_A+5\beta_0\Big]\,,
\notag\\[2mm]
 r_{gg}&=-8C_A n_f+4C_F n_f
+4C_A^2\left({17} -{\pi^2}-6\zeta_3\right)\,,
\end{align}
and
\begin{align}
\omega_{gg}(\tau)&=\frac{16}3 C_A n_f\left\{\bar\tau\left(\ln\bar\tau-\frac{50}3\right) +23\right\}
+ 40 C_F n_f \left\{\bar\tau\left(\ln\bar\tau+\frac25\right)+2\tau\right\}
\notag\\
&\quad +16 C_A^2\Biggl\{
2(1+2\tau )\left[\Li_2(\tau)+\frac{\pi^2}6\right]+3(1+\tau)\ln^2\bar\tau
-\frac{3-4\tau-14\tau^2+3\tau^3}{6\tau}\ln\bar\tau
\notag\\
&\quad
-\frac1{36}\big(457+1007\tau\big)
 \Biggr\}\,.
\end{align}
We close this discussion with a remark on the so-called reciprocity symmetry of the invariant kernels~\cite{Basso:2006nk}. This
symmetry arises, technically, from the observation~\cite{Basso:2006nk} that invariant kernels can in general be presented in terms
of the quadratic Casimir operator of the collinear conformal group.
{ %
For an operator with conformal spin $j$ (in our
case $j=n+2$) and its anomalous dimension given by $\gamma(j)=f(j+\frac12\gamma(j))$, the asymptotic expansion of the function
$f(j)$ for large $j$ consists of terms invariant under $j\to 1-j$. It was shown in Ref.~\cite{Braun:2017cih} that the function
$f(j)$ defined by this equation gives the eigenvalues} of the invariant kernel, $\mathbb{H}^{\text{inv}}$, hence the eigenvalues of
$\mathbb{H}^{\text{inv}}$ should have (and indeed they have) the corresponding invariance property. The reciprocity relation has
been checked on many examples for the situations where only one operator exists for a given conformal spin. In the situation that
there are two and or more operators of the same conformal spin, the reciprocity cannot be expected in general for the off-diagonal
elements of the
{%
anomalous dimension matrix, because} they depend explicitly on the  assumed  normalization for the
operators~\cite{Basso:2006nk}. Since the evolution kernels for the LROs do not depend on the basis of local operators, one should
expect, however, that in this representation the reciprocity holds for the off-diagonal elements (kernels) as well. Indeed, one can
verify that the eigenvalues of the off-diagonal invariant kernels, Eqs.~\eqref{GQ-spectrum} and \eqref{QG-spectrum} are invariant
under
$j\to 1-j$ {%
with $j=n+2$.
For the invariant kernels themselves this condition implies an expansion $\sum_k h_k(\tau) \ln^k\tau$
for $\tau\to 0$, where $h_k(\tau)$ is an analytic function in the vicinity of $\tau=0$.}

\section{Anomalous dimension matrix for local operators}\label{sect:LRtoLocal}

For a comparison with Ref.~\cite{Belitsky:1998gc} and also for the application to the scale-dependence of the meson
{%
light-cone distribution amplitudes,} it is desirable to have the results also in a different form,
as an anomalous dimension matrix for local operators in the Gegenbauer basis~\eqref{Gegenbauer}.

The coefficient functions $\Psi^{(\chi)}_{nk}(z_1,z_2)$ in the expansion of the quark and gluon LROs over this basis~\eqref{Gegenbauer}
\begin{align}\label{LRO-def-ren-2}
[\mathcal{O}_\chi(z_1,z_2)]=\sum_{nk} \Psi^{(\chi)}_{nk}(z_1,z_2) [\mathcal{O}_{\chi,nk}(0)]\,.
\end{align}
can be obtained by the repeated application of the canonical $S^{(0)}_+$ generator to the coefficient function of the
{%
highest-weight} operator $k=n$,
\begin{align}\label{coeff-Psi-functions}
\Psi^{(q)}_{nk}(z_1,z_2) \,=\,  \omega_{nk} \,(S_+^{q,(0)})^{k-n}\, z_{12}^n \,,
&&
\Psi^{(g)}_{nk}(z_1,z_2) \,=\, \omega_{nk} \,(S_+^{g,(0)})^{k-n}\, z_{12}^{n-1}\,,
\end{align}
where
\begin{align}
 \omega_{nk} =  2 \frac{2 n + 3}{(k-n)!} \frac{\Gamma(n + 2) }{\Gamma(n + k + 4)}\,.
\end{align}
These polynomials are mutually orthogonal and form a complete set of functions w.r.t. the canonical $SL(2)$ scalar product (see,
e.g., {
Ref.}~\cite{Braun:2011dg})
\begin{align} \label{ortho}
  \langle \Psi^{(q)}_{nk} | \Psi^{(q)}_{n'k'} \rangle_{j=1}
=\delta_{kk'}\delta_{nn'} \,\omega_{nk} \rho_n^{-1} \,,
&&
  \langle \Psi^{(g)}_{nk} | \Psi^{(g)}_{n'k'} \rangle_{j=3/2}
=\delta_{kk'}\delta_{nn'} \, \frac{4}{n(n+3)} \omega_{nk} \rho_n^{-1} \,,
\end{align}
where
\begin{align}
 \rho_n = \frac12{(n+1)(n+2)!}\,.
\end{align}
The local operators~\eqref{Gegenbauer} can be obtained by the projection of the LROs
on the corresponding coefficient function
\begin{align}
  [ \mathcal O_{q,nk}]  = \frac{ \langle \Psi^{(q)}_{nk} | [\mathcal O_q(z_1,z_2)] \rangle_{j=1}}
                               {\langle \Psi^{(q)}_{nk} | \Psi^{(q)}_{nk} \rangle_{j=1}}
                         \,,
&& 
  [ \mathcal O_{g,nk}]  = \frac{\langle \Psi^{(g)}_{nk} | [\mathcal O_g(z_1,z_2)] \rangle_{j=3/2}}
                               {\langle \Psi^{(g)}_{nk} | \Psi^{(g)}_{nk} \rangle_{j=3/2}}
                         \,.
\end{align}
In order to translate the action of the evolution kernel $\mathbb H$ in the LRO representation in Eq.~\eqref{eq:H2loop}
in terms of local operators we adopt the following dictionary:
\begin{align}\label{eq:localTOlightray}
  \boldsymbol{A}^{qq}_{nn'} & =
  \frac{  \langle \Psi^{(q)}_{nk} | \mathbb  A_{qq} | \Psi_{n'k}^{(q)} \rangle_{(j=1)} }
       {\langle \Psi^{(q)}_{nk} | \Psi^{(q)}_{nk} \rangle_{j=1}} \,,
  &&
  \boldsymbol{A}_{nn'}^{qg} =
  \frac{ \langle \Psi^{(q)}_{nk} | \mathbb  A_{qg} | \Psi_{n'k}^{(g)} \rangle_{(j=1)} }
       {\langle \Psi^{(q)}_{nk} | \Psi^{(q)}_{nk} \rangle_{j=1}} \,,
       \nonumber\\
  \boldsymbol{A}^{gq}_{nn'} &=
  \frac{  \langle \Psi^{(g)}_{nk} | \mathbb  A_{gq} | \Psi_{n'k}^{(q)} \rangle_{(j=3/2)} }
       {\langle \Psi^{(g)}_{nk} | \Psi^{(g)}_{nk} \rangle_{j=3/2}} \,,
  &&
  \boldsymbol{A}_{nn'}^{gg} =
  \frac{ \langle \Psi^{(g)}_{nk} | \mathbb  A_{gg} | \Psi_{n'k}^{(g)} \rangle_{(j=3/2)} }
       {\langle \Psi^{(g)}_{nk} | \Psi^{(g)}_{nk} \rangle_{j=3/2}}\,,
\end{align}
where $\mathbb A \in \{ \mathbb H, \mathbb X, \mathbb T\}$.
Note that for all operators from this set $[\mathbb A, S_-] = 0$ and, therefore, we can
choose $k=n$ in Eq.~\eqref{eq:localTOlightray} for convenience without loss of generality.
Moreover, it is worth to be mentioned that for any invariant operator
$\mathbb A_{\rm inv}^{\chi\chi'}$
the matrix $\boldsymbol{A}_{\rm inv}^{\chi\chi'}$ is diagonal:
\begin{align}
  \boldsymbol{A}_{\rm inv,nn'}^{\chi \chi'} = \delta_{nn'} A_{\rm inv,n}^{\chi \chi'}\,.
\end{align}
Making use of these properties we find the following results for the local matrix elements of the
$\mathbb X$ operators~\eqref{eq:Xoneloop}
\begin{align}
  \boldsymbol{X}^{(1)\chi\chi'} _{nm} =\delta_{nm}X^{(1)}_{\chi\chi'}(n)
  - \vartheta_{nm}\frac{w^{(1)\chi\chi'}_{nm}}{a(n,m)}\,,
\end{align}
with the eigenvalues $X^{(1)}_{\chi\chi'}(n)$ given in {%
Eqs.~\eqref{eq:X1gq} and \eqref{eq:X-spectrum}},
the discrete step function defined as
\begin{align}
  \vartheta_{nm} =
  \begin{cases}
    1 \text{ if } n-m>0 \text{ and even}\,,
    \\
    0 \text{ else}\,,
  \end{cases}
\end{align}
and the following matrices:
\begin{align}
  w^{(1)qq}_{jk} & =
  4 C_F (2 k + 3) a(j, k) \biggl[
  \frac{A_{j, k} - S_1(j + 1)}{(k + 1) (k + 2)}
  + 2 \frac{A_{j,k}}{a(j,k) } \biggr]\,, \nonumber \\
  w^{(1)qg}_{jk} & = 0\,, \nonumber \\
  w^{(1)gq}_{jk} & = 4 C_F (2 k + 3) \frac{(j-k)(j+k+3)}{(k+1)(k+2) }\,, \nonumber \\
  w^{(1)gg}_{jk} & = 4 C_A (2 k + 3) \biggl\{
  2 A_{j,k} + \big[A_{j,k}- S_1(j + 1) \big]\left[\frac{\Gamma(j+4)\Gamma(k)}{\Gamma(k+4)\Gamma(j)}-1\right]
\notag\\&\qquad
				    +2 (j-k)(j+k+3) \frac{\Gamma(k)}{\Gamma(k+4)} \biggr\}\,,
\end{align}
where
\begin{align}
  A_{j, k} & =  S_1\left(\frac{j + k + 2}2\right) - S_1\left(\frac{j - k - 2}2\right)
  + 2 S_1(j - k - 1) - S_1(j + 1)\,,  \nonumber \\
  a(j,k) & =  (j-k)(j+k+3)\,.
\end{align}
For the operator $\mathbb T^{(1)}$ defined in Eq.~\eqref{eq:Toneloop} we obtain
\begin{align}
  \boldsymbol{T}_{nm}^{(1)\chi\chi'} = \delta_{nm}\partial_n h_{\chi\chi'}^{(1)}(n)
  + \vartheta_{nm}\frac{2(2m+3)}{a(n,m)} \big(h^{(1)}_{\chi\chi'}(n) - h^{(1)}_{\chi\chi'}(m)\big) \,,
\end{align}
with the one-loop anomalous dimensions in {%
Eqs.}~\eqref{h1qg+gq}, \eqref{h1qq+gg}, \eqref{h1derivative}.

This yields the following representation for the anomalous dimension matrix $\boldsymbol{H}_{nm}^{(2)\chi\chi'}$
of the local operators
\begin{align}
\boldsymbol{H}_{nm}^{(2)PS} &= \delta_{nm}h_{PS}^{\rm inv} (n)
    + h_{qg}^{(1)}(n) \boldsymbol{X}^{(1)gq}_{nm} + \frac12 \boldsymbol{T}^{(1)qg}_{nm}h_{gq}^{(1)}(m)\,,
\nonumber \\
\boldsymbol{H}_{nm}^{(2)NS} &= \delta_{nm}h_{NS}^{\rm inv} (n)
    + \big[h_{qq}^{(1)}(n) - h_{qq}^{(1)}(m) \big] \boldsymbol{X}^{(1)qq}_{nm}
        +\boldsymbol{T}^{(1)qq}_{nm}\big[\beta_0 + \frac12h_{qq}^{(1)}(m) \big]\,,
\nonumber\\
\boldsymbol{H}_{nm}^{(2)qg} &= \delta_{nm}h_{qg}^{\rm inv} (n)  + h_{qg}^{(1)}(n) \boldsymbol{X}^{(1)gg}_{nm}
    - \boldsymbol{X}^{(1)qq}_{nm}h_{qg}^{(1)}(m)
        +  \boldsymbol{T}^{(1)qg}_{nm}\big[\beta_0 + \frac12h_{gg}^{(1)}(m) \big]
            +  \frac12 \boldsymbol{T}^{(1)qq}_{nm}h_{qg}^{(1)}(m) \,,
\nonumber \\
\boldsymbol{H}_{nm}^{(2)gq} &= \delta_{nm}h_{gq}^{\rm inv} (n)  + h_{gq}^{(1)}(n) \boldsymbol{X}^{(1)qq}_{nm}
  +h_{gg}^{(1)}(n) \boldsymbol{X}^{(1)gq}_{nm} - \boldsymbol{X}^{(1)gq}_{nm}h_{qq}^{(1)}(m)
  - \boldsymbol{X}^{(1)gg}_{nm} h_{gq}^{(1)}(m)
\nonumber \\
&\quad +  \boldsymbol{T}^{(1)gq}_{nm}\big[\beta_0 + \frac12h_{qq}^{(1)}(m) \big]
  +  \frac12 \boldsymbol{T}^{(1)gg}_{nm}h_{gq}^{(1)}(m) \,,
\nonumber \\
    \boldsymbol{H}_{nm}^{(2)gg} &= \delta_{nm}h_{gg}^{\rm inv} (n)
        + \big[h_{gg}^{(1)}(n) - h_{gg}^{(1)}(m) \big] \boldsymbol{X}^{(1)gg}_{nm}
        - \boldsymbol{X}^{(1)gq}_{nm}h_{qg}^{(1)}(m)
\nonumber \\
&\quad
    +  \boldsymbol{T}^{(1)gg}_{nm}\big[\beta_0 + \frac12h_{gg}^{(1)}(m) \big]
                +  \frac12 \boldsymbol{T}^{(1)gq}_{nm}h_{qg}^{(1)}(m) \,.
\end{align}
The eigenvalues $h_{\chi\chi'}^{\rm inv} (n)$ of the two-loop invariant kernels can be found in
Eqs.~\eqref{PS-spectrum},~\eqref{GQ-spectrum},~\eqref{QG-spectrum} and~\eqref{GG-spectrum}.

The result in this form can directly be compared to the two-loop anomalous dimension matrix
calculated in Ref.~\cite{Belitsky:1998gc}.
The diagonal elements $\boldsymbol{H}^{(2)\chi\chi'}_{nn}$ for each channel reproduce the well-known eigenvalues
$h_{\chi\chi'}^{(2)}(n)$ of the evolution kernel by construction,
{%
Eqs.}~\eqref{eq:pstwoloopspectrum}, \eqref{eq:twoloopspectrum},
and in both calculations coincide (up to a different normalization) with the anomalous dimensions from Ref.~\cite{Vogt:2004mw}. For
the non-diagonal elements note that our definition of the gluonic local operator in Eq.~\eqref{Gegenbauer} differs from the one
used in Ref.~\cite{Belitsky:1998gc} by a factor 6, i.e.
\begin{align}
  \mathcal O_{g,nk}^{\rm this\,work} = 6 \mathcal O_{g,nk}^{\text{{\color{blue}
  Ref.}\cite{Belitsky:1998gc}}},
\end{align}
and also that the anomalous dimensions in Ref.~\cite{Belitsky:1998gc} are written as an expansion in $\alpha_s/(2\pi)$.
Taking these differences into account we find a perfect agreement.

\section{Evolution kernels in the momentum fraction representation}

Evolution equations in the momentum fraction representation can be derived straightforwardly from the kernels in the coordinate
representation. Let $f(x_1,x_2)$ be a Fourier transform of the position-space distribution $f(z_1,z_2)$ defined as
\begin{align}
f(z_1,z_2)=\int dx_1 dx_2 e^{-iz_1x_1-iz_2 x_2} f(x_1,x_2)\,.
\end{align}
The general expression for an evolution kernel in the momentum fraction representation takes the form
\begin{align}\label{komega-1}
(\mathcal{H} f)(x_1,x_2)=\int \mathcal{D}u \mathcal{H}(x_1,x_2|u_1,u_2) f(u_1,u_2)\,,
\end{align}
where $\mathcal{D}u=du_1 du_2 \delta(x_1+x_2-u_1-u_2)$ {%
and} the $\delta$-function is due to the momentum conservation.
 Following the decomposition in Eq.~\eqref{three-terms-expansion} we split the  kernel $\mathcal{H}$  into three parts
\begin{align}
\mathcal{H}= C_0 \II +  \widehat{\mathcal{H}}_{\vartheta} + \mathcal{H}_{\omega}\,,
\end{align}
where $\widehat{\mathcal{H}}_{\vartheta}$, $\mathcal{H}_{\omega}$ originate from the contributions
{%
which have the form in the coordinate space (LRO) representation}
\begin{align}
(\widehat{\mathcal{H}}_{\vartheta} f)(z_1,z_2) &=
    \int_0^1 d\alpha\frac{\bar\alpha}{\alpha} \vartheta(\bar\alpha) \Big(2f(z_1,z_2)-f(z_{12}^\alpha,z_2)-f(z_1,z_{21}^\alpha)\Big)\,,
\notag\\
({\mathcal{H}}_{\omega} f)(z_1,z_2) &=
    \int_0^1 d\alpha\int_0^{\bar\alpha} d\beta \,\omega(\alpha,\beta) f(z_{12}^\alpha,z_{21}^\beta)\,.
\end{align}
It is easy to show that the corresponding kernels in the momentum fraction representation are given by the following expressions:
\begin{align}
(\widehat{\mathcal{H}}_{\vartheta} f)(x_1,x_2)
&=\int \mathcal{D}{u}\left\{
{\Theta(x_1,u_1-x_1)}\frac{\vartheta(x_1/u_1)}{u_1-x_1}
+ (x_1,u_1\leftrightarrow x_2,u_2)
\right\}
\big(f(x_1,x_2)-
f(u_1,u_2)\big)\,,
\notag\\
({\mathcal{H}}_{\omega} f)(x_1,x_2) &=\int \mathcal{D}{u}\biggl\{ {\Theta(x_1,-x_2,u_1-x_1)} A_\omega(x_i,x'_i)
\notag\\
&\quad +{\Theta(x_1,x_2,u_1-x_1)}B_\omega(x_i,x'_i) +
{\Theta(x_1,x_2,u_2-x_2)}C_\omega(x_i,x'_i)\biggr\} f(u_1,u_2)\,.
\end{align}
The $\Theta$ function is defined as follows
\begin{align}\label{multitheta}
\Theta(a_1,\ldots,a_n)=\prod_{k=1}^n \theta(a_i)-\prod_{k=1}^n \theta(-a_i)\,,
\end{align}
and functions $ A_\omega, B_\omega, C_\omega$ are given by the following integrals
\begin{align}\label{A-int}
A_\omega(x_i,u_i) &=\frac1{x'_1}\int_0^{(x_1-x'_1)/x'_2} d\beta\,
\omega(\alpha_x,\beta) =-\frac1{x'_2}\int_0^{(x_2-x'_2)/x'_1} d\alpha\,
\omega(\alpha,\beta_x)\,,
\notag\\
B_\omega(x_i,u_i) &= \frac{1}{x'_1}\,\int_0^{x_1/(x_1+x_2)} d\beta  \,\omega(\alpha_x,\beta)\,,
\notag\\
C_\omega(x_i,u_i) &= \frac{1}{x'_2}\,\int_0^{x_2/(x_1+x_2)} d\alpha  \,\omega(\alpha,\beta_x)\,,
\end{align}
where it is implied that $x_1+x_2=u_1+u_2$ and
\begin{align}
\alpha_x&=(x_2-\bar\beta u_2)/{u_1}, &&\beta_x=(x_1-\bar\alpha u_1)/{u_2}\,.
\end{align}
The kernel $\omega(\alpha,\beta)$ is a single-valued function in the simplex $0<\alpha+\beta<1$.
It is easy to see that if the variables $x_i,u_i$ belong to the regions determined by the corresponding $\Theta$-functions, then
the arguments $\alpha_x,\beta_x$ {%
lie} inside the simplex and the corresponding integrals are unambiguously defined.
In the present case, $\omega(\alpha,\beta)$ is symmetric under the interchange of {%
 its} two arguments
$\omega(\alpha,\beta)=\omega(\beta,\alpha)$ (this is not always the case, see {%
Ref.}~\cite{Braun:2009mi}),
and as a consequence $C_\omega(x_1,x_2,u_1,u_2) = B(x_2,x_1,u_2,u_1)$. Provided
that the functions $B_\omega, C_\omega$ are continued in an appropriate way  beyond their analyticity domain one gets
$A_\omega = B_\omega-C_\omega$.
Note also that all these kernels are effectively functions of two variables,
$x=x_1/(x_1+x_2)$ and $y=u_1/(x_1+x_2)$, e.g. $A_\omega= a(x,y)/(x_1+x_2)$, etc.

All integrals that arise in the transition from the position to the momentum fraction representation can be taken analytically in
terms of elementary functions and $\text{Li}_2$ polylogarithms. The resulting expressions are collected in
Ref.~\cite{Belitsky:1999hf}  and are implemented in the NLO evolution FORTRAN code by Freund and
McDermott~\cite{Freund:2001rk,Freund:2001hd}~\footnote{\url{http://durpdg.dur.ac.uk/hepdata/dvcs.html}}.  We have checked
numerically that our results for the kernels in the momentum fraction representations agree with the kernels implemented in this
code for all color structures. In this way we can confirm the results of Ref.~\cite{Belitsky:1999hf}, where these kernels are given
in analytic form~\footnote{{%
See footnote 17 in Ref.~\cite{Freund:2001hd}.}}.

\section{Summary}\label{sect:summary}

{%
We have presented a re-derivation} of the two-loop flavor-singlet evolution
equations for the leading-twist operators in off-forward kinematics, based on using conformal symmetry
of QCD in non-integer $d-2\epsilon$ space-time dimensions at the critical point.
This case is more complicated as compared to the flavor-nonsinglet evolution, both, technically and conceptually,
due to potentially dangerous contributions of gauge-noninvariant
operators. Our analysis is based on {%
studies}
of conformal Ward identities in Ref.~\cite{Braun:2018mxm}, where it is proven that such extra terms
do not contribute to correlation functions with gauge-invariant operators.

The results in this work are given in the position-space or light-ray operator representation. This form has some technical
advantages and {%
can also be interesting in applications to lattice QCD} calculations. Expanding our results in powers of the field
separation we reproduce the results for the mixing matrices for flavor-singlet local operators derived in
Ref.~\cite{Belitsky:1998gc}. The evolution kernels in momentum fraction space can be obtained from our expressions by simple
integration. We have verified numerically that the resulting kernels agree with the kernels that are implemented in the NLO GPD
evolution code by Freund and McDermott~\cite{Freund:2001rk,Freund:2001hd} which is based on the analytic expressions derived in
Ref.~\cite{Belitsky:1999hf}.

{%
To summarize, the two-loop evolution equations for generalized parton distributions
have now been derived independently by two groups, and perfect agreement is found.
In view of the projected high statistics of the relevant experiments at the JLAB 12 GeV upgrade and, in future, the EIC,
it is necessary to implement and use these results now in NLO analyses of
deeply-virtual Compton scattering and similar reactions.}

\section*{Acknowledgments}
We are grateful to V.~Guzey for providing us with the source code for the NLO GPD evolution by Freund and
McDermott~\cite{Freund:2001rk,Freund:2001hd}.
This work was supported by the DFG with the grants MO~1801/1-3 (A.M.) and SFB/TRR~55 (M.S.).


\appendix
\section*{Appendices}
\addcontentsline{toc}{section}{Appendices}

\renewcommand{\theequation}{\Alph{section}.\arabic{equation}}
\renewcommand{\thesection}{{\Alph{section}}}
\renewcommand{\thetable}{\Alph{table}}
\setcounter{section}{0}
\setcounter{table}{0}

\section{Scale and conformal transformations}\label{App:A}

Scale ($D$) and conformal  ($K$) field transformations  for the fundamental fields take the generic form
\begin{align}\label{}
\delta_D \Phi(x)
    &=D_{\Delta_\Phi}(x) \Phi(x)
    \big(x\partial_x+\Delta_\Phi\big) \Phi(x)\,,
\notag\\
\delta_{K^\mu} \Phi(x)
    &=K^\mu_{\Delta_\Phi}(x) \Phi(x)
    \big(2x^\mu(x\partial)-x^2\partial^\mu +2\Delta_\Phi x^\mu -2x_\nu \Sigma^{\mu\nu}\big) \Phi(x)\,,
\end{align}
in particular
\begin{align}\label{}
K_\mu q(x)
    &=\big(2x_\mu(x\partial)-x^2\partial_\mu +2\Delta_q\, x_\mu\big)\, q(x)+\frac12[\gamma_\mu,\slashed{x}]q(x)\,,
\notag\\
K_\mu \bar q(x)
    &=\big(2x_\mu(x\partial)-x^2\partial_\mu +2\Delta_q\, x_\mu\big)\, \bar q(x)-\bar q(x)\frac12[\gamma_\mu,\slashed{x}]\,,
\notag\\
K_\mu c(x)
    &=\big(2x_\mu(x\partial)-x^2\partial_\mu +2\Delta_c\, x_\mu\big)\, c(x)\,,
\notag\\
K_\mu \bar c(x)
    &=\big(2x_\mu(x\partial)-x^2\partial_\mu +2\Delta_{\bar c}\, x_\mu\big)\, \bar c(x)\,,
\notag\\
K_\mu A_\rho(x)
    &=\big(2x_\mu(x\partial)-x^2\partial_\mu +2\Delta_A\, x_\mu\big)\, A_\rho(x) + 2g_{\mu\rho} (x A)-2x_\rho A_\mu(x)\,,
\end{align}
where $\Delta_\Phi=\dim \Phi$ are the canonical dimensions of the fields.
It is convenient {%
to choose} them in $4-2\epsilon$ dimensions to be exactly the same as in the four-dimensional theory,
\begin{align}
\Delta_A=1, && \Delta_{q}=\Delta_{\bar q}=\frac32, && \Delta_c=0, &&\Delta_{\bar c}=2.
\end{align}
For this choice the gluon strength tensor $F_{\sigma\rho}$ transforms in a covariant way 
\begin{flalign}\label{F-transform}
K_\mu F_{\sigma\rho} &=\Big(2x_\mu(x\partial)-x^2\partial_\mu +4 x_\mu\Big)F_{\sigma\rho}
+2\Big(g_{\mu\rho} x^\nu F_{\sigma\nu}+g_{\mu\sigma} x^\nu F_{\nu\rho} -x_\rho F_{\sigma\mu}-x_\sigma F_{\mu\rho}\Big)\,, &
\end{flalign}
and the covariant derivative of the ghost field $D_\nu c$ transform as a vector field,
\begin{flalign}\label{}
K_\mu D_\rho c(x) &=\big(2x_\mu(x\partial)-x^2\partial_\mu +2x_\mu\big) D_\rho c(x)
+2\big(g_{\mu\rho} (x D)-x_\rho D_\mu\Big)c(x)\,. &
\end{flalign}

The variation  of the different parts of the QCD action under the special conformal transformation takes the form
\begin{subequations}
\begin{align}
\label{Kqq}
\delta_K\int d^dx \bar q \slashed{D} q &=
    4\epsilon \int d^dx \left(x^\mu \bar q \slashed{D} q~+~\frac12 \bar q \gamma_\mu q\right)\,,
\\
\label{F2}
\delta_K\int d^dx \frac14 F^2 &=
    4\epsilon \int d^dx \,x^\mu\frac14 F^2\,,
\\
\label{KGFt}
\delta_K\int d^dx\frac1{2\xi}(\partial A)^2 &=
    -\frac1\xi\int d^dx\Big(-2\epsilon\, x^\mu (\partial A)^2+2(d-2) A^\mu (\partial A)\Big)\,,
\\
\label{Sghost}
\delta_K\int d^dx\Big(-\bar c\partial_\mu D^\mu c\Big) & =
    4\epsilon \int d^dx x^\mu\Big(-\bar c\partial_\mu D^\mu c\Big) +2(d-2)\int d^dx \, \bar c D^\mu c\,.
\end{align}
\end{subequations}
Note that the ghost and the gauge-fixing terms break conformal symmetry explicitly even in $d=4$ dimensions.
Summing up all contributions one obtains
\begin{flalign}
\label{DS}
\delta_D S & =\int d^dx \,2\epsilon\mathcal{L}(x)\,,   \\
\label{KS}
\delta_{K^\mu} S &= \int d^dx\,\left(4\epsilon\, x^\mu\left( \mathcal{L}(x) -\frac12 \partial^\rho
\mathcal{J}_\rho(x)\right)-2(d-2)\partial^\rho \mathcal{B}_{\rho}(x)\right)\,.
\end{flalign}
Here $\mathcal{J}_{\rho}(x)=\bar q(x) \gamma_\rho q(x)$ is the conserved flavor-singlet current and
\begin{flalign}
\label{Bmu}
\mathcal{B}_\mu=\bar c D_\mu c -\frac1\xi A_\mu (\partial A)
\end{flalign}
is a BRST variation, $\mathcal{B}_\mu =\delta_{\rm BRST}(\bar c A_\mu)$.

\section{One-loop conformal anomaly}\label{App:B}

Our analysis  follows closely the lines of Ref.~\cite{Braun:2016qlg} so that in this Appendix we concentrate mainly on specific
problems that arise for the flavor-singlet operators.

Consider~\cite{Braun:2016qlg} the correlation function of two (renormalized) LROs stretched in different directions
\begin{align}\label{FInt1}
\boldsymbol{G}_{\chi\chi'}(x;z,w)=\VEV{[\mathcal{O}_\chi^{(n)}](0,z)\,[\mathcal{O}_{\chi'}^{(\bar n)}](x,w)}
=\VEV{\widehat{\mathcal{O}}_\chi^{(n)}(0,z)\,\widehat{\mathcal{O}}_{\chi'}^{(\bar n)}(x,w)}\,,
\end{align}
where $z=\{z_1,z_2\}$, $w=\{w_1,w_2\}$ and $n$, $\bar n$ are two auxiliary light-like vectors, $n^2=\bar n^2=0$.
In what follows we assume that $(n\bar n)=1$ and $(x\cdot n)=(x\cdot \bar n)=0$.
Starting from the representation in terms of local conformal operators it can be shown, see Ref.~\cite{Braun:2016qlg},  that the
correlation function~\eqref{FInt1} satisfies the constraint
\begin{align}\label{CWI-explicit}
\sum_{\sigma=q,g} S_{+,\chi \sigma }^{(z)}\boldsymbol{G}_{\sigma \chi'}(x;z,w)=\frac12 x^2
(\bar n\partial_x)\boldsymbol{G}_{\chi\chi'}(x;z,w)\,,
\end{align}
where the operator $S_{+,\sigma \chi}$ is the generator of special conformal transformations defined in {%
Eq.}~\eqref{def-Splus}.
The explicit expression for $S_{+,\sigma \chi}$ can be found from the  analysis of the conformal Ward identity (CWI) for
the correlation function~\eqref{FInt1}. Making the corresponding change of variables in the path integral
representation for this correlation function one obtains
\begin{align}\label{SCWI}
\VEV{\delta \widehat{\mathcal{O}}_\chi^{(n)}(0,z)\,\widehat{\mathcal{O}}_{\chi'}^{(\bar n)}(x,w)}
+\VEV{\widehat{\mathcal{O}}_\chi^{(n)}(0,z)\,\delta \widehat{\mathcal{O}}_{\chi'}^{(\bar n)}(x,w)}
=\VEV{\delta S_R\, \widehat{\mathcal{O}}_\chi^{(n)}(0,z)\,\widehat{\mathcal{O}}_{\chi'}^{(\bar n)}(x,w)}\,,
\end{align}
where $\delta$ stands for the conformal variation along the $\bar n$ direction, $\delta =\bar n^\mu \delta_{K^\mu}$. The second
term on the l.h.s. is solely responsible for the r.h.s. of Eq.~\eqref{CWI-explicit} while the two others contribute to the l.h.s.
of this relation. The first of them, {the} variation of the LRO operator, takes the form
\begin{align}
\delta \widehat{\mathcal{O}}_\chi^{(n)}(0,z)  &= Z\delta { \mathcal{O}}_\chi^{(n)}(0,z)=
2(n\bar n)Z S_{\chi,+}^{(0)}{\mathcal{O}}_\chi^{(n)}(0,z)
=
2(n\bar n) Z S_{\chi,+}^{(0)}Z^{-1}\widehat{\mathcal{O}}_\chi^{(n)}(0,z)\,.
\end{align}
The renormalization factor $Z$ in this equation is an integral operator in $z_1,z_2$ and it does not commute with $S_{\chi,+}^{(0)}$.
However, since $S_{\chi,+}^{(0)}$ does not have any $\epsilon$ dependence,
\begin{align}
Z S_{\chi,+}^{(0)}Z^{-1}= S_{\chi,+}^{(0)} + \text{ singular terms in $1/\epsilon$}\,.
\end{align}
Since all singular terms must cancel in the final result we can drop them.

The second term that contributes to the l.h.s. of Eq.~\eqref{CWI-explicit} is due to the conformal
variation of action as given in Eq.~\eqref{KS}. As the first step, it
has to be re-expanded  in terms of renormalized operators. In Landau gauge that we will assume from now on, the
corresponding expression takes the form
\begin{align}\label{LQCD-LG-prime-1}
{2\epsilon}\mathcal{L}^\prime_R(x)& =
-\frac{\beta(a)}{a}\left[\mathcal{L}^{YM}\right]-(\gamma_q-\epsilon) \Omega_{q\bar q}
+2\gamma_c\Big(\Omega_A - \partial_\mu [\mathcal{B}^\mu]\Big)-2(\gamma_c-\epsilon)\Omega_{\bar c}\,,
\end{align}
where $\mathcal{B}^\mu$ is defined in {Eq.}~\eqref{Bmu},
   $\mathcal{L}'_R(x)=\mathcal{L}_R(x)-\frac12 \, \partial^\rho [\mathcal{J}_\rho](x) $,
$\Omega_\Phi =\Phi {\delta S_R}/{\delta \Phi}$, $\Omega_{q\bar q}=\Omega_q+\Omega_{\bar q}$.
Note that the anti-ghost EOM term does not contribute to the
correlation function in question since the LROs $\widehat{\mathcal{O}}_\chi$ do not contain
(anti)ghost fields. The terms $\Omega_A$ and
$\partial_\mu[\mathcal{B}^\mu]$ contain {%
contributions of the gauge parameter proportional to $1/\xi$,}
which, however, cancel each other.
For nonzero $\xi$ the correlation function of $\mathcal{B}_\mu$ with any gauge invariant functional
$\mathcal{F}$ vanishes, implying that
\begin{align}
\lim_{\xi\to0}\VEV{\Big(\Omega_A - \partial_\mu [\mathcal{B}^\mu]\Big)(x)\, \mathcal{F}}_\xi=
\VEV{ A(x)\frac{\delta\mathcal{F}}{\delta A(x)}}_{\xi=0}\,.
\end{align}
Integrating by parts { the} contributions of the EOM terms in $\vev{\delta S_R\,
\widehat{\mathcal{O}}_\chi^{(n)}(0,z)\,\widehat{\mathcal{O}}_{\chi'}^{(\bar n)}(x,w)}$ we can write, e.g., for the first operator
insertion,
\begin{flalign}\label{Lxi0}
{2\epsilon}\mathcal{L}^\prime_R(x)\widehat{\mathcal{O}}_\chi^{(n)}(0,z) &=
\left(-\frac{\beta(a)}{a}\left[\mathcal{L}^{YM}(x)\right]
+(\epsilon-\gamma_q)\mathcal{D}_{q\bar q}(x) +2\gamma_c \mathcal{D}_A(x)\right) \widehat{\mathcal{O}}_\chi^{(n)}(0,z)\,,
\end{flalign}
where we introduce the notation $\mathcal{D}_\Phi(x)=\Phi(x)\frac{\delta }{\delta \Phi(x)}$,
 $\mathcal{D}_{q\bar q}(x) = \mathcal{D}_{q}(x) + \mathcal{D}_{\bar q}(x)$.

{%
 Let us consider first the quark operator, $\widehat{\mathcal{O}}_q(0,z)$. In order to streamline} the notation, hereafter, we do
not display the dependence on the auxiliary vector $n$ and the space-time coordinates, if this is clear from the context. Our goal
is to calculate the one-loop correction to the generators. It can be shown that there are no BRST or EOM counterterms to the quark
LRO at one loop, i.e.
    $\widehat{\mathcal{O}}_q(z)= Z_{qq}\mathcal{O}_q(z) +Z_{qg}\mathcal{O}_g(z) +O(a^2) $.
It is also obvious that the one-loop correction to the quark part of the generator, $S_{qq}$,  is the same as in the
non-singlet case. We consider therefore only the off-diagonal entry $S_{qg}$, i.e. we are looking for the contributions of the
gluon LRO  $\widehat{\mathcal{O}}_g$ on the r.h.s. of Eq.~\eqref{Lxi0}.

First, note that the term $2\gamma_c \mathcal{D}_A(x)\widehat{\mathcal{O}}_q(z)$
gives rise to a $O(a^2)$ contribution only and can be omitted. Second, there are no gluon pair counterterms for the product
$\left[\mathcal{L}^{YM}(x)\right]\widehat{\mathcal{O}}_q(z)$  at one loop.
Thus, the only relevant contribution comes from the quark EOM term. We write
\allowdisplaybreaks[0]{
\begin{align}
 \int d^dx\, (\bar n x) \mathcal{D}_{q\bar q}(x) \widehat{\mathcal{O}}_q(z) &=
 \int d^dx\, (\bar n x)\mathcal{D}_{q\bar q}(x) \Big(Z_{qq}\mathcal{O}_q(z) +Z_{qg}\mathcal{O}_g(z)\Big)
 =
 Z_{qq} \int d^dx\, (\bar n x)\mathcal{D}_{q\bar q}(x)\mathcal{O}_q(z) \notag\\
    &=(n\bar n) Z_{qq}(z_1+z_2) \mathcal{O}_q(z)
        =(n\bar n)Z_{qq}(z_1+z_2)\Big(
         Z^{-1} _{qq} \widehat{\mathcal{O}}_q(z)+
             Z^{-1}_{qg} \widehat{\mathcal{O}}_g(z)\Big)\,.
\end{align}
}

\noindent
The gluon  contribution of interest corresponds to the last term in this expression. Taking into account that the renormalization
factor is related to the evolution kernel as
$ Z^{-1}_{qg}=-\frac a{2\epsilon} \mathbb{H}_{qg}^{(1)}+O(a^2)$ and multiplying this relation by $\epsilon-\gamma_q$,
one can bring this contribution to the form
\begin{align}
-(n\bar n) \frac12 (z_1+z_2) a_*
\mathbb{H}_{qg}^{(1)}\widehat{\mathcal{O}}_g(z)+ O(a^2)\,.
\end{align}
This result implies that $\Delta^{(1)}_{qg}=0$ in agreement with {Ref.}~\cite{Belitsky:1998gc}.

Next, consider the $S_{gq}$ generator. To this end we again omit the $\mathcal{D}_A$ term in {Eq.}~\eqref{Lxi0} since it is
$O(a^2)$. The quark EOM contribution in {Eq.}~\eqref{Lxi0} can be handled as above yielding
\begin{align}\label{Dgq1}
(n\bar n) \frac12  a_*
\mathbb{H}_{gq}^{(1)}(z_1+z_2)\widehat{\mathcal{O}}_q(z)+ O(a^2)\,.
\end{align}
In addition we have to consider the contribution from the product  $\left[\mathcal{L}^{YM}(x)\right]\widehat{\mathcal{O}}_g(z)$.
This product is divergent when $x\to 0$ so that one has to add and subtract the corresponding pair counterterms.
The counterterm $\sim \mathcal{O}_q$ corresponds to the Feynman diagram shown in Fig.~\ref{fig:FFtoQQ}
%
%
\begin{figure}[t]
\centerline{\includegraphics[width=0.2\textwidth]{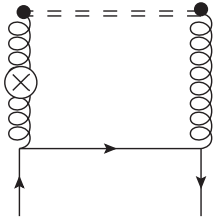}}
\caption{One-loop quantum correction  to $S_{gq}$.}
\label{fig:FFtoQQ}
\end{figure}
%
%
where the crossed circle stands for the insertion of the vertex
\begin{align}
2  \int d^dx (\bar n x) L^{YM}_{\text{kin}}(x)= -\int d^dx (\bar n x) A^a_\mu(x)
\partial^2 A^a_\mu(x)+ \mathcal{O}((\partial A))\,.
\end{align}
The terms involving $(\partial A)$ vanish since the propagators are transverse in Landau gauge.
The gluon line with the crossed-circle insertion (modified propagator) can be written in the form
\begin{align}
\mathcal{G}_{\mu\nu}(x-y) &=G_{\mu\nu}(x-y)((\bar n x) + (\bar n y)) + H_{\mu \nu}(x-y)\,,
\end{align}
where $G_{\mu\nu}$ is the usual gluon propagator in Landau gauge and
\begin{align}
H_{\mu \nu}(x)=i\int\frac{d^dp}{(2\pi)^d}\frac{\bar n^\mu p^\nu-\bar n^\nu p^\mu}{p^4}e^{-ipx}\,.
\end{align}
Taking into account the contribution of the crossing-symmetric diagram one obtains (divergent part only)~\footnote{The
contributions due to $H_{\mu\nu}$  vanish  in the sum of all diagrams. }
\begin{align}
-\frac1{2\epsilon}(n\bar n) a \big\{\mathbb{H}^{(1)}_{gq}, z_1+z_2\big\}_+ \mathcal{O}_{q}(z)
+\frac1\epsilon (n\bar n)   2C_F a \int_\alpha \Big(\mathcal{O}_{q}(z_{12}^\alpha,z_2)-\mathcal{O}_{q}(z_1,z_{21}^\alpha)\Big)\,,
\end{align}
where $\{\ldots,\ldots\}_+$ stands for the anti-commutator. Multiplying this expression by {$(-\beta(a)/2a)$} and adding
{Eq.}~\eqref{Dgq1} we obtain the result for $\Delta^{(1)}_{gq}$ given in Eq.~\eqref{Delta-1-loop}.

The calculation of the pure gluon contribution, $\Delta_{gg}$, is more involved. It was shown in \cite{Braun:2018mxm} that the
r.h.s. of Eq.~\eqref{Lxi0} can be represented as sum of the fully renormalized product $[F_{\mu\nu}^2(x) \mathcal{O}_g(z)]$ which
enters with the coefficient $\beta(a)/a$, contributions of gauge-invariant operators and gauge-noninvariant contributions that are
BRST variations of nonlocal operators or nonlocal EOM operators which drop out of all correlation functions with gauge-invariant
operators, hence also of {Eq.}~\eqref{SCWI}. Nevertheless, we need to know the structure of the gauge-noninvariant terms in order
to correctly separate the gauge-invariant contributions. Consider first the BRST and EOM counterterms to the gluon LRO
$\mathcal{O}_g$. Since they should be twist-two operators, there are two possible structures:
\begin{align}
B_k=\delta_{\rm BRST}( (\partial_+ \bar c)(z_1) A_+(z_2)\ldots A_+(z_k)),
&& E_k=\frac{\delta S}{\delta A_+(z_1)} A_+(z_2)\ldots A_+(z_k)\,.
\end{align}
Here we tacitly assume that the color indices are contracted in some way. {%
Both operators} contain singular terms in $\xi$
that have to cancel in the sum. Therefore they can appear only in the combination,
$B_k(z_1,\ldots,z_k)-E_k(z_1,\ldots,z_k)$. The lowest, $k=2$, counterterm can be restored by calculating the one-loop
$F_{+\mu}F_{+\mu} \mapsto \bar c c $ diagram. The result reads
\begin{align}
\label{B-E}
(B-E)_2=-\frac{a C_A}{\epsilon}  \int_0^{1}d\alpha\int_0^{1} d\beta \,
\left\{ \delta_{\rm BRST}\Big([\partial_+ \bar c^a](z_{12}^\alpha) A_+^a(z_{21}^\beta)\Big)
-A_+^a (z_{21}^\beta) \frac{\delta S}{\delta A_+^a} (z_{12}^\alpha)
\right\}\,.
\end{align}
The BRST variation of this operator is given by a sum of EOM contributions which, in agreement with the general statement,
do not contain the anti-ghost EOM.
{%
The contributions for $k=3,4,\ldots$} can be neglected to our accuracy as they are higher order in the coupling.

Next we notice that for the calculation of $\Delta_{gg}$ to one-loop accuracy one can omit the quark EOM term
$\sim \mathcal{D}_{q\bar q}$ in {Eq.}~\eqref{Lxi0}, and the remaining contribution $\sim \mathcal{D}_A(x)$ can be written as
\begin{align}\label{O23g}
2\gamma_c\int d^dx (\bar n x)\mathcal{D}_A(x)\mathcal{O}_g(z)=
2\gamma_c\Big((n\bar n)(z_1+z_2) \mathcal{O}_g(z)+
 \mathcal{O}_{2,g}(z)+ (n\bar n)\mathcal{O}_{3,g}(z)\Big)\,.
\end{align}
Here
\begin{flalign}\label{Delta23g}
 \mathcal{O}_{2g}(z)&=\widetilde A^a_\mu(z_1) F^a_{+\mu}(z_2)
 + (z_1\leftrightarrow z_2)\,, \hskip 5mm \widetilde A_\mu =(n\bar n)A_\mu-\bar n^\mu A_+\,, &
 \\
\label{Delta233g}
\mathcal{O}_{3g}(z)&=g f^{abc}  
\Big[z_1 A^b_+(z_1) A^c_\mu (z_1) F^a_{+\mu}(z_2)\! +\! (z_1\!\leftrightarrow\! z_2)
+z_{12}F^c_{+\mu}(z_1)\!\! \int_\alpha z_{21}^\alpha A_+^b(z_{21}^\alpha)   F^a_{+\mu}(z_2)\Big]\,, &
\end{flalign}
where the gauge links (in the adjoint representation) are tacitly implied. Since {Eq.}~\eqref{O23g} is multiplied by the ghost
anomalous dimension
$2\gamma_c$
the contribution of $\mathcal{O}_{3g}$ is higher order in the coupling and can be omitted to the $O(a)$ accuracy. Thus the first
term in {Eq.}~\eqref{Lxi0} can be presented in the form
\begin{align}
\left[\mathcal{L}^{YM}(x)\right]\widehat{\mathcal O}_g(z)=\left[\mathcal{L}^{YM}(x)\right][{\mathcal O}_g(z)]-
\left[\mathcal{L}^{YM}(x)\right](B-E)_2(z) +\ldots \,,
\end{align}
where $(B-E)_2$ is given in {Eq.}~\eqref{B-E} and the ellipses stand for $k>2$ (gauge-{noninvariant}) counterterms which contribute
only starting from $O(a^2)$. The term with the BRST variation $B_2$ also drops out from the correlation function at order $O(a)$.
The remaining EOM term can be rewritten (inside the correlation function) with the required accuracy as
\begin{align}
\label{bad}
-\frac{a C_A}{\epsilon} \int d^d x (\bar n x)
    \int_0^{1}d\alpha\int_0^{1} d\beta A^a_+(z_{21}^\beta) \frac{\delta F_{\mu\nu}^2(x)}{\delta A^a_+(z_{12}^\alpha)}\,.
\end{align}
This is the only contribution from the BRST and EOM counterterms at this order.

Note that the operator in {Eq.}~\eqref{bad} is not a BRST variation so that, according to the general statement of
Ref.~\cite{Braun:2018mxm}, this contribution must be complemented by contributions of other diagrams to produce a BRST variation or
EOM operator in the sum, which is not easy to see on a diagrammatic level. For our purposes it is, however, sufficient to note that
all such ``unwanted'' contributions contain either $A_+$ or $A_-$ fields, and can be avoided if one looks for the contributions of
transverse gluons.

 In order to calculate the pair counterterm to  the operator product $\left[\mathcal{L}^{YM}(x)\right][{\mathcal O}_g(z)]$ it
is convenient to make some rearrangements. Namely, we rewrite
\begin{align}\label{LtoL}
[\mathcal{L}^{YM}]\mapsto\Big [\mathcal{L}^{YM}-\frac12 \left(\Omega_A-\partial^\mu B_\mu\right)\Big]+
\frac12 \Big[\Omega_A-\partial^\mu B_\mu\Big]\,.
\end{align}
The integral of the first piece can be written as (up to $(\partial A)$ contributions)
\begin{align}
-\frac12 \int d^d x (\bar n x)\Big [\mathcal{L}^{YM}-\frac12 \left(\Omega_A-\partial^\mu B_\mu\right)\Big] &=
-\frac12 \int d^d x (\bar n x)\Big [\mathcal{L}^{YM}_{3A}+2\mathcal{L}^{YM}_{4A}-\partial\bar c\partial c +\Omega_{\bar c}\Big]
\notag\\&
\mapsto -\frac12 \int d^d x (\bar n x)\Big [\mathcal{L}^{YM}_{3A}+2\mathcal{L}^{YM}_{4A}\Big]\,,
\end{align}
where $\mathcal{L}_{3,4}^{YM}$ stand for the three- and four-gluon vertices.
The ghost terms can be omitted as they do not contribute to the
gauge-invariant counterterms. In the second piece, the BRST operator does not contribute to the correlation function
and the
$\Omega_A$-term gives
\begin{align}\label{EOM-A}
    \frac12 \int d^dx (\bar n x) \mathcal{D}_A(x)[\mathcal{O}_g(z)]\,.
\end{align}
All these contributions are multiplied by ${(-\beta(a)/2a)}=2(\epsilon -\gamma_g)$.

The main contribution to the CWI is due to the pair counterterms to
\begin{align}\label{Snx}
-\int d^d x (\bar n x)\Big [\mathcal{L}^{YM}_{3A}+2\mathcal{L}^{YM}_{4A}\Big] [{\mathcal O}_g(z)]\,.
\end{align}
The corresponding diagrams are shown in Fig.~\ref{fig:FFtoFF} (adding the symmetric contributions is implied).
%
\begin{figure}[t]
\centerline{\includegraphics[width=0.87\textwidth]{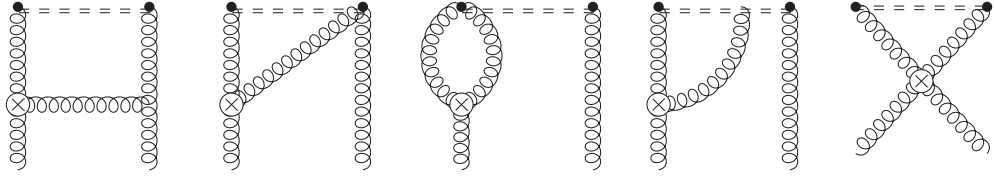}}
\caption{Gluon diagrams generated by the contribution in Eq.~\eqref{Snx}. }
\label{fig:FFtoFF}
\end{figure}
%
They can be related to the counterterm diagrams for the gluon LRO with some extra terms.

Since the four-gluon vertex does not contain derivatives, the counterterm corresponding to the
last diagram in Fig.~\ref{fig:FFtoFF} can be rewritten as
\begin{align}
\label{four}
\int d^dy A_\mu^{a}(y) A_\nu^b(y)\widebar G_{\mu\nu}^{ab}(y,z_1,z_2)= \int d^d x (\bar n x) \mathcal{D}_A(x)
\int d^dy  A_\mu^{a}(y) A_\nu^b(y)G_{\mu\nu}^{ab}(y,z_1,z_2)\,,
\end{align}
where $G_{\mu\nu}^{ab}(y,z_1,z_2)$ is {the} usual QCD diagram with the four-gluon vertex and $\widebar G_{\mu\nu}^{ab}(y,z_1,z_2)$
is the diagram  with the modified vertex,
$ \widebar G_{\mu\nu}^{ab}(y,z_1,z_2)\equiv 2 (\bar n y)G_{\mu\nu}^{ab}(y,z_1,z_2) $.

For the rest of the contributions in Fig.~\ref{fig:FFtoFF} that involve a three-gluon vertex (which contains a derivative),
this relation has to be modified as follows. Let $y_1$ be the coordinate of the modified vertex
(insertion of $(\bar n\cdot y_1)$). Then
\begin{align}
\label{three}
\int d^dy_1 d^d y_2 A(y_1) A(y_2 )\widebar G(y_1,y_2,z_1,z_2)&=%
\int d^dy_2 A(y_2)\! \int d^d x (\bar n x) \mathcal{D}_A(x)
\!\int d^dy_1  A(y_1) G(y_1,y_2,z_1,z_2)
\notag\\
&\quad
+\int d^dy_1 d^d y_2 A(y_1) A(y_2 )\widetilde G_1(y_1,y_2,,z_1,z_2)\,.
\end{align}
where we suppressed  color and Lorentz indices on the gauge field.
In this expression, as above, $G(y_1,y_2,z_1,z_2)$ is the diagram with the standard QCD  three-gluon vertex,
$\widebar G(y_1,y_2,z_1,z_2) = \ (\bar n y_1) G(y_1,y_2,z_1,z_2)$ and $\widetilde G_1(y_1,y_2,,z_1,z_2)$
is the Feynman diagram obtained from $G(y_1,y_2,z_1,z_2)$ by the replacement of the triple vertex $V(q,r, k)$
at the position $y_1$ by the new vertex $\widetilde V(q,r, k)$ defined as
\begin{align}\label{triGM}
\widetilde V(q,r, k) =  -i (\bar n,\partial_q) V(q,r, k)\bigr|_{q=0}\,.
\end{align}
Using these relations the divergent parts of the contributions in Fig.~\ref{fig:FFtoFF} can be presented in the form
\begin{align}\label{diag-S}
\int d^dx (\bar n x)\mathcal{D}_A(x) \Big(\text{CT}\big( \mathcal{O}_g\big)\Big) + \sum \widetilde D_i
-\sum D^\prime_i\,.
\end{align}
Here $\text{CT}(O_g)$ stands for the counterterms to the operator $\mathcal{O}_g$,
$\widetilde D_i$ are the first four diagrams in Fig.~\ref{fig:FFtoFF} (with the triple-gluon vertex)
with the crossed vertex replaced by {Eq.}~\eqref{triGM} and $D^\prime_i$ are the diagrams shown in Fig.~\ref{fig:FF-1} (upper row),
where the big black circle stands for the vertices generated by the operator
$
(n\bar n)A_\mu(z_2)-\bar n^\mu A_+(z_2)+
(n\bar n) z_2 F_{+\mu}(z_2)\,
$.
%
\begin{figure}[t]
\centerline{\includegraphics[width=0.57\textwidth]{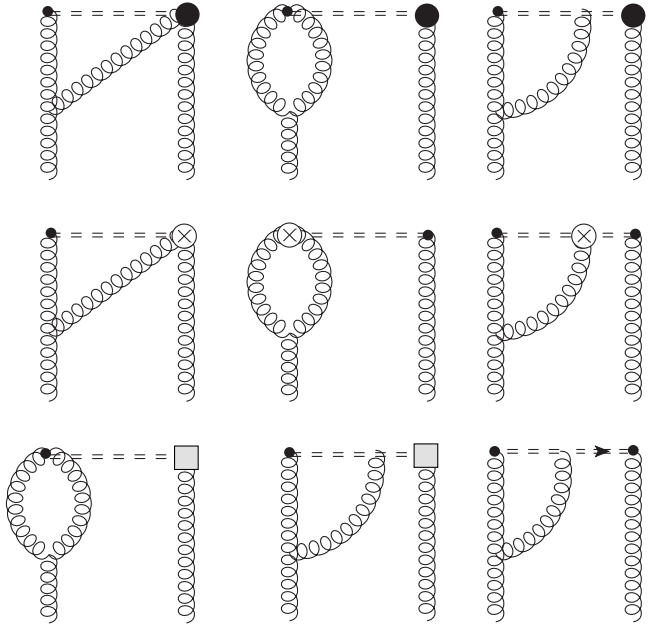}}
\caption{
Upper row: $D^\prime_i$ contribution to Eq.~\eqref{diag-S}.
Middle row: Counterterms to the operator $\mathcal{O}_{3g}$ \eqref{Delta233g}.
Lower row:
Sum of the diagrams in the upper and the middle rows. The shaded box stands for the $(n\bar n)A_\mu-\bar n^\mu A_+$ vertex
and the arrow for $(z_2-w)(n\bar n)$.
}
\label{fig:FF-1}
\end{figure}
The first contribution in Eq.~\eqref{diag-S} absorbs all terms with  the derivative $\mathcal{D}_A$ in {Eqs.}~\eqref{three} and
\eqref{four} but also generates additional contributions, the diagrams  $D^\prime_i$ shown in the upper row in Fig.~\ref{fig:FF-1},
that have, therefore, to be subtracted.

We recall that {Eq.}~\eqref{diag-S} has to be multiplied by factor ${(-\beta(a)/2a)}=\epsilon-\gamma_g$ and also the contribution
due to {Eq.}~\eqref{EOM-A} has to be added, which reads
\begin{align}\label{EOM-AA}
(\epsilon-\gamma_g)\int d^dx (\bar n x) \mathcal{D}_A(x)[\mathcal{O}_g](z)\,.
\end{align}
The first term in {Eq.}~\eqref{diag-S} together with {Eq.}~\eqref{EOM-AA} gives
\begin{align}
\label{g23}
(\epsilon-\gamma_g)\int d^dx (\bar n x) \mathcal{D}_A(x)\mathcal{O}_g(z)=
(\epsilon-\gamma_g)\Big((n\bar n)(z_1+z_2) \mathcal{O}_g(z)+
 \mathcal{O}_{2,g}(z)+ (n\bar n)\mathcal{O}_{3,g}(z)\Big)\,,
\end{align}
where the operators $\mathcal{O}_{2,g}(z)$ and  $\mathcal{O}_{3,g}(z)$ are defined in {Eqs.}~\eqref{Delta23g} and
\eqref{Delta233g}, respectively. This expression has still to be rewritten as the sum of renormalized operators minus the
corresponding counterterms. Since $\gamma_g(a_*)=\epsilon$ the contributions of renormalized operators vanish at the critical
point. For the first term we can write
\begin{align}
(\gamma_g - \epsilon)(n\bar n)(z_1+z_2) \text{CT}(\mathcal{O}_g(z))=
-\frac a2 (n\bar n) (z_1+z_2) \Big(\mathbb{H}^{(1)}_{gg}-2\gamma_A\Big)
 \mathcal{O}_g(z)+
O(a^2) +\text{BRST} +\text{EOM}\,.
\end{align}
Combining this expression with {Eq.}~\eqref{O23g} and taking into account that in Landau gauge $2\gamma_c=-\gamma_A-\gamma_g$ one
obtains the usual contribution to the generator $S_+$
\begin{align}
- (n\bar n) a(z_1+z_2) \left(-\gamma_g +\frac12\mathbb{H}^{(1)}_{gg} \right)\mathcal{O}_g(z)\underset{a\to a_*}{\longrightarrow}
- (n\bar n)(z_1+z_2)a_* \left(-\epsilon +\frac12\mathbb{H}^{(1)}_{gg} \right)\mathcal{O}_g(z)\,.
\end{align}
The counterterms to the operator $\mathcal{O}_{3g}$ in {Eq.}~\eqref{g23} come from the diagrams shown in the middle row in
Fig.~\ref{fig:FF-1}. The first of them cancels with the first diagram in the upper row. The remaining two diagrams in the upper and
the middle row do not cancel completely but their sum can be simplified to the diagrams shown in lowest row. Strictly speaking,
there is also a contribution of the form $(z_1-z_2)\mathcal{O}_g(z_1,z_2)$ that cancels in the sum with the symmetric diagram
($z_1\leftrightarrow z_2$). Finally, the counterterm diagrams for the operator $\mathcal{O}_{2g}$ are shown in Fig.~\ref{fig:O2g}.
%
\begin{figure}[t]
\centerline{\includegraphics[width=0.97\textwidth]{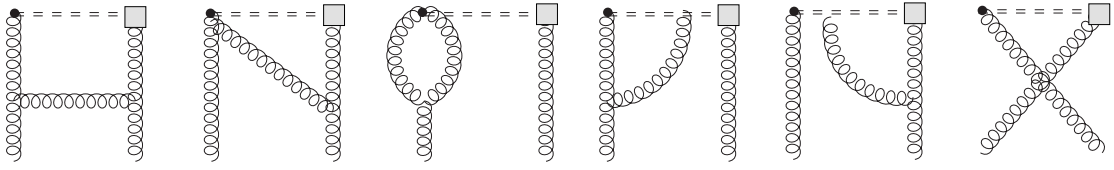}}
\caption{The counterterm diagrams for the operator $\mathcal{O}_{2g}$. The shaded box stands for the
vertex~$(n\bar n)A_\mu-\bar n^\mu A_+$.}
\label{fig:O2g}
\end{figure}
%
The  third and the fourth diagram in Fig.~\ref{fig:O2g} cancel with the first and the second diagrams in the lowest row in
Fig.~\ref{fig:FF-1}, respectively. The remaining ones give rise to the conformal anomaly $\Delta_{gg}$. { These are:}
\begin{itemize}
\item {The} first four diagrams in Fig.~\ref{fig:FFtoFF} with the triple gluon vertex~\eqref{triGM}.
\item {The} last diagram in the lowest row in Fig.~\ref{fig:FF-1}.
\item {The} first, second, fifth and sixth diagram in Fig.~\ref{fig:O2g}.
\end{itemize}
Technically, it is convenient to combine the first two diagrams in Fig.~\ref{fig:FFtoFF} with the first two diagrams in
Fig.~\ref{fig:O2g} as there are strong cancellations. The calculation of all other diagrams is rather straightforward. Note, that
the final answer contains not only the term $F_{\mu+} F_{\mu+}$ which we are interested in, but also gauge non-invariant
contributions. These terms contain the field $A_+$ or $A_-$ or both of them and must be part of the BRST and EOM operators which
can appear on the r.h.s. of {Eq.~\eqref{Lxi0}. We did not keep track} of such contributions. But we have checked that the
gauge-noninvariant terms, $(n\bar n) A_\mu(z_1n) F_{+\mu}(z_2 n)$, which are neither BRST nor EOM operators, cancel out in the sum
of all diagrams.
 The resulting expression for the
conformal anomaly is presented in Eq.~\eqref{Delta-1-loop}. It has a very simply form and coincides with the result of Belitsky and
M\"uller~\cite{Belitsky:1998gc}, {that is} the ${}^{\chi\chi'}\mathcal{K}^\omega$ kernels in their notation.

\section{RG identities in Landau gauge}\label{app:Landau}
The analysis of the Ward identities in this work is based on the re-expansion of the renormalized Lagrangian
$2\epsilon \mathcal{L}_R$ in terms of renormalized operators.
The corresponding general expression in arbitrary covariant gauge reads~\cite{Spiridonov:1984br,Belitsky:1998gc,Braun:2016qlg}
\begin{align}\label{LQCDR}
{2\epsilon}\mathcal{L}'_R&=-\frac{\beta(a)}{a}\left[\mathcal{L}^{YM}+\mathcal{L}^{gf}\right]-(\gamma_q-\epsilon) \Omega_{q\bar q}
-(\gamma_A+\gamma_g)\Omega_A-(\gamma_c-2\epsilon)\Omega_{\bar c} -\gamma_c\Omega_c+2\gamma_A [\mathcal{L}^{gf}]
\notag\\
&\quad
+z_b(g,\xi) \partial_\mu [\mathcal{B}^\mu]+z_c(g,\xi)\partial_\mu [\Omega^\mu]\,.
\mathrm{}
\end{align}
Here $\mathcal{L}'_R=\mathcal{L}_R-\frac12 \partial^\rho(\bar q_0\gamma_\mu q_0)$,
$\gamma_\Phi$ are the anomalous dimensions of the fields,
$\Omega_\Phi(x)=\Phi(x){\delta
S_R}/{\delta\Phi(x)}$ and
$\Omega_\mu= \bar c D_\mu c-\partial_\mu \bar c c$ is a conserved current,
$\partial_\mu [\Omega^\mu]=\Omega_c-\Omega_{\bar c}$. The coefficients $z_b(g,\xi)$ and $z_c(g,\xi)$ are at this
stage unknown functions.

The gauge fixing term,
 $[\mathcal{L}^{gf}]=\frac1{2\xi} (\partial A)^2$, can be written as a combination of BRST and EOM operators,
\begin{align}
\frac1\xi (\partial A)^2=-\mathcal{B}-\Omega_{\bar c}\,, &&    \mathcal{B}=\delta_{\rm BRST}\big(\bar c^a\,(\partial A^a)\big)\,.
\end{align}
{An} arbitrary diagram with an insertion of this operator vanishes in Landau gauge. Hence, this contribution can safely be omitted.
Next, whereas all counterterms are polynomials in the gauge parameter $\xi$, the terms $\Omega_A$ and
$\partial_\mu[\mathcal{B}^\mu]$ on the r.h.s. of {Eq.}~\eqref{LQCDR} contain contributions
$\sim 1/\xi$ which have to cancel in the sum.
This implies that $z_g(g,\xi)=\gamma_A+\gamma_g +O(\xi)$. The last coefficient $z_c(g,\xi)$ can be fixed in Landau gauge by
observing that in this gauge {the} QCD action has an additional discrete symmetry corresponding to the interchange of ghost and
antighost fields, $c\to -\bar c$, $\bar c \to c$. Indeed it is easy to check that
\begin{align}
\mathcal{L}(-\bar c, c)-\mathcal{L}(c,\bar c)=\partial^\mu \Omega_\mu -g f^{abc} \bar c^a (\partial A)^b c^c\,.
\end{align}
Since the last term can be omitted in Landau gauge,  the action is invariant with respect to this transformation.
Clearly Eq.~\eqref{LQCDR} should respect this property. Taking into account that
$\Omega_c+\Omega_{\bar c}$ is invariant under this transformation, $\Omega^\mu\mapsto -\Omega^\mu$, and
\begin{align}
\Omega_A - \partial_\mu [\mathcal{B}^\mu]\to \Omega_A - \partial_\mu [\mathcal{B}^\mu]+\partial^\mu \Omega_\mu
\,,
\end{align}
one obtains $z_c(g,\xi\to 0)=\gamma_c=-(\gamma_A+\gamma_g)/2$. Using these expressions we obtain
the following result
\begingroup
 \allowdisplaybreaks
\begin{align}\label{LQCD-LG-prime-2}
{2\epsilon}\mathcal{L}'_R & =
-\frac{\beta(a)}{a}\left[\mathcal{L}^{YM}\right]-(\gamma_q-\epsilon) \Omega_{q\bar q}
+2\gamma_c\Big(\Omega_A - \partial_\mu [\mathcal{B}^\mu]\Big)-2(\gamma_c-\epsilon)\Omega_{\bar c}
\notag\\
&=
-\frac{\beta(a)}{a}\left[\mathcal{L}\right]-(\gamma_q+\gamma_g) \Omega_{q\bar q}
-(\gamma_A+\gamma_g)\Big(\Omega_A - \partial_\mu [\mathcal{B}^\mu]\Big)
-(\gamma_c+\gamma_g)\Big(\Omega_{\bar c}+\Omega_c-\partial^\mu\Omega_\mu\Big)
\notag\\
&=-\frac{\beta(a)}{a}\left[\mathcal{L}\right] -\sum_\Phi (\gamma_\Phi+\gamma_g) \Omega_\phi
+(\gamma_A+\gamma_g) \partial_\mu [\mathcal{B}^\mu] + (\gamma_c+\gamma_g)\partial^\mu\Omega_\mu\,,
\end{align}
\endgroup
which holds in Landau gauge. { Note, that here we do not keep} the gauge fixing term $\mathcal{L}^{gf}$, since it does not
contribute in this gauge.

\section{Evolution kernels vs. anomalous dimensions}\label{App:D}
In this Appendix we collect the expressions for the invariant kernels that appear as building blocks in two-loop evolution
equations and { the anomalous dimensions.} The anomalous dimension  $\gamma(j)$ corresponding to the kernel $h(\tau)$ is defined as
follows:
\begin{align}
\gamma(j)=\int_{\alpha\beta} h\left(\tau\right)(1-\alpha-\beta)^{j-1}\equiv \mathcal{M}\Big[h(\tau)\Big]\,.
\end{align}
The inverse relation reads~\cite{Braun:2014vba}
\begin{align}
h(\tau)=\frac1{2\pi i}\int_C dj \,(2j+1)\gamma(j)\, P_j\left(\frac{1+\tau}{1-\tau}\right)\,,
\end{align}
where it is assumed that all poles in anomalous dimensions are located to the left of the
integration contour which is along the imaginary axis.
Here $P_j(z)$ is the  Legendre function,
\begin{align}
P_j\left(\frac{1+\tau}{1-\tau}\right)=(1-\tau)^{-j} {}_2F_1(-j,-j, 1, \tau)\,.
\end{align}
Below we list the relevant { expressions for the anomalous dimensions $\gamma(j)$ (left column) and the corresponding ones for the
kernels $h(\tau)$ (right column):}
\begin{flalign*}
&\frac1{j(j+1)}  &&  \mathcal{M}\Big[1\Big] &\\
&\frac1{j^2(j+1)^2}  &&  \mathcal{M}\Big[-\ln\bar\tau\Big]\\
&\frac1{(j+2)(j-1)} &&  \mathcal{M}\Big[(1+\tau)/\bar\tau\Big]\\
&\frac1{(j+2)^2(j-1)^2} &&   \mathcal{M}\left[\frac1{3\bar\tau}\Big(2\tau-(1+\tau)\ln\bar\tau\Big)\right]\\
&\frac1{(j+2)^3(j-1)^3}&&\mathcal{M}
\Biggl[
\frac1{9\bar\tau}\Biggl((1+\tau)\left(\Li_2(\tau)+\frac12\ln^2\bar\tau+\frac13\ln\bar\tau\right)
\notag\\
&&&\quad
-2\ln\bar\tau-\frac83\tau \Biggr)\Biggr]
\\
&\frac{1}{j(j+1)} S_1(j)&&  \mathcal{M}\left[-\frac12\ln\tau\right]
\\
&\frac{S_1(j)-1}{(j-1)j(j+1)(j+2)} &&  \mathcal{M}\left[-\frac12 \frac\tau{\bar\tau} \ln\tau\right]
\\
& (-1)^j\Big[S_{-2}(j)+{\pi^2}/{12}\Big] &&  \mathcal{M}\left[\frac12\bar\tau\right]
\\
&\frac{(-1)^j}{j(j+1)}
\left[S_{-2}(j)+{\pi^2}/{12}\right] && \mathcal{M}\left[ \frac12 \Li_2(\tau)\right]
\\
&(-1)^{j}\frac{S_{-2}(j)+\pi^2/12}{(j-1)j(j+1)(j+2)} &&  \mathcal{M}\left[\frac1{2\bar\tau}\Big(\Li_2(\tau)-\bar\tau\ln\bar\tau-2\tau\Big)\right]
\\
&S_3(j)-\zeta_3 && \mathcal{M}\left[ \frac 12 \frac{\bar\tau}{\tau}\ln\bar\tau\right]
\\
&(-1)^{j}\Big[S_{-3}(j)-2S_{1,-2}(j)
-S_1(j)\frac{\pi^2}{6}+\frac12\zeta_3\Big] && \mathcal{M}\left[ -\frac12\bar\tau\ln\bar\tau\right]
\end{flalign*}
%
%

%

\begin{thebibliography}{10}

\bibitem{Accardi:2012qut}
A.~Accardi et~al., \emph{{Electron Ion Collider: The Next QCD Frontier}},
  \href{https://doi.org/10.1140/epja/i2016-16268-9}{\emph{Eur. Phys. J.}
  {\bfseries A52} (2016) 268},
  [\href{https://arxiv.org/abs/1212.1701}{{\ttfamily 1212.1701}}].

\bibitem{Accardi:2016ndt}
A.~Accardi et~al., \emph{{A Critical Appraisal and Evaluation of Modern PDFs}},
  \href{https://doi.org/10.1140/epjc/s10052-016-4285-4}{\emph{Eur. Phys. J.}
  {\bfseries C76} (2016) 471},
  [\href{https://arxiv.org/abs/1603.08906}{{\ttfamily 1603.08906}}].

\bibitem{Belitsky:1999hf}
A.~V. Belitsky, A.~Freund and D.~M{\"u}ller, \emph{{Evolution kernels of skewed
  parton distributions: Method and two loop results}},
  \href{https://doi.org/10.1016/S0550-3213(00)00012-2}{\emph{Nucl. Phys.}
  {\bfseries B574} (2000) 347--406},
  [\href{https://arxiv.org/abs/hep-ph/9912379}{{\ttfamily hep-ph/9912379}}].

\bibitem{Mueller:1991gd}
D.~M{\"u}ller, \emph{{Constraints for anomalous dimensions of local light cone
  operators in phi**3 in six-dimensions theory}},
  \href{https://doi.org/10.1007/BF01555504}{\emph{Z. Phys.} {\bfseries C49}
  (1991) 293--300}.

\bibitem{Mueller:1993hg}
D.~M{\"u}ller, \emph{{Conformal constraints and the evolution of the nonsinglet
  meson distribution amplitude}},
  \href{https://doi.org/10.1103/PhysRevD.49.2525}{\emph{Phys. Rev.} {\bfseries
  D49} (1994) 2525--2535}.

\bibitem{Mueller:1997ak}
D.~M{\"u}ller, \emph{{Restricted conformal invariance in QCD and its predictive
  power for virtual two photon processes}},
  \href{https://doi.org/10.1103/PhysRevD.58.054005}{\emph{Phys. Rev.}
  {\bfseries D58} (1998) 054005},
  [\href{https://arxiv.org/abs/hep-ph/9704406}{{\ttfamily hep-ph/9704406}}].

\bibitem{Belitsky:1998gc}
A.~V. Belitsky and D.~M{\"u}ller, \emph{{Broken conformal invariance and
  spectrum of anomalous dimensions in QCD}},
  \href{https://doi.org/10.1016/S0550-3213(98)00677-4}{\emph{Nucl. Phys.}
  {\bfseries B537} (1999) 397--442},
  [\href{https://arxiv.org/abs/hep-ph/9804379}{{\ttfamily hep-ph/9804379}}].

\bibitem{Belitsky:1998vj}
A.~V. Belitsky and D.~M{\"u}ller, \emph{{Next-to-leading order evolution of
  twist-2 conformal operators: The Abelian case}},
  \href{https://doi.org/10.1016/S0550-3213(98)00310-1}{\emph{Nucl. Phys.}
  {\bfseries B527} (1998) 207--234},
  [\href{https://arxiv.org/abs/hep-ph/9802411}{{\ttfamily hep-ph/9802411}}].

\bibitem{Braun:2013tva}
V.~M. Braun and A.~N. Manashov, \emph{{Evolution equations beyond one loop from
  conformal symmetry}},
  \href{https://doi.org/10.1140/epjc/s10052-013-2544-1}{\emph{Eur. Phys. J.}
  {\bfseries C73} (2013) 2544},
  [\href{https://arxiv.org/abs/1306.5644}{{\ttfamily 1306.5644}}].

\bibitem{Braun:2007wv}
V.~Braun and D.~M{\"u}ller, \emph{{Exclusive processes in position space and
  the pion distribution amplitude}},
  \href{https://doi.org/10.1140/epjc/s10052-008-0608-4}{\emph{Eur. Phys. J.}
  {\bfseries C55} (2008) 349--361},
  [\href{https://arxiv.org/abs/0709.1348}{{\ttfamily 0709.1348}}].

\bibitem{Ji:2013dva}
X.~Ji, \emph{{Parton Physics on a Euclidean Lattice}},
  \href{https://doi.org/10.1103/PhysRevLett.110.262002}{\emph{Phys. Rev. Lett.}
  {\bfseries 110} (2013) 262002},
  [\href{https://arxiv.org/abs/1305.1539}{{\ttfamily 1305.1539}}].

\bibitem{Ma:2017pxb}
Y.-Q. Ma and J.-W. Qiu, \emph{{Exploring Partonic Structure of Hadrons Using ab
  initio Lattice QCD Calculations}},
  \href{https://doi.org/10.1103/PhysRevLett.120.022003}{\emph{Phys. Rev. Lett.}
  {\bfseries 120} (2018) 022003},
  [\href{https://arxiv.org/abs/1709.03018}{{\ttfamily 1709.03018}}].

\bibitem{Braun:2014vba}
V.~M. Braun and A.~N. Manashov, \emph{{Two-loop evolution equations for
  light-ray operators}},
  \href{https://doi.org/10.1016/j.physletb.2014.05.037}{\emph{Phys. Lett.}
  {\bfseries B734} (2014) 137--143},
  [\href{https://arxiv.org/abs/1404.0863}{{\ttfamily 1404.0863}}].

\bibitem{Braun:2016qlg}
V.~M. Braun, A.~N. Manashov, S.~Moch and M.~Strohmaier, \emph{{Two-loop
  conformal generators for leading-twist operators in QCD}},
  \href{https://doi.org/10.1007/JHEP03(2016)142}{\emph{JHEP} {\bfseries 03}
  (2016) 142}, [\href{https://arxiv.org/abs/1601.05937}{{\ttfamily
  1601.05937}}].

\bibitem{Braun:2017cih}
V.~M. Braun, A.~N. Manashov, S.~Moch and M.~Strohmaier, \emph{{Three-loop
  evolution equation for flavor-nonsinglet operators in off-forward
  kinematics}}, \href{https://doi.org/10.1007/JHEP06(2017)037}{\emph{JHEP}
  {\bfseries 06} (2017) 037},
  [\href{https://arxiv.org/abs/1703.09532}{{\ttfamily 1703.09532}}].

\bibitem{Freund:2001rk}
A.~Freund and M.~F. McDermott, \emph{{A Next-to-leading order QCD analysis of
  deeply virtual Compton scattering amplitudes}},
  \href{https://doi.org/10.1103/PhysRevD.65.074008}{\emph{Phys. Rev.}
  {\bfseries D65} (2002) 074008},
  [\href{https://arxiv.org/abs/hep-ph/0106319}{{\ttfamily hep-ph/0106319}}].

\bibitem{Freund:2001hd}
A.~Freund and M.~McDermott, \emph{{A Detailed next-to-leading order QCD
  analysis of deeply virtual Compton scattering observables}},
  \href{https://doi.org/10.1007/s100520200928}{\emph{Eur. Phys. J.} {\bfseries
  C23} (2002) 651--674},
  [\href{https://arxiv.org/abs/hep-ph/0111472}{{\ttfamily hep-ph/0111472}}].

\bibitem{Balitsky:1987bk}
I.~I. Balitsky and V.~M. Braun, \emph{{Evolution Equations for QCD String
  Operators}}, \href{https://doi.org/10.1016/0550-3213(89)90168-5}{\emph{Nucl.
  Phys.} {\bfseries B311} (1989) 541--584}.

\bibitem{Vogt:2004mw}
A.~Vogt, S.~Moch and J.~A.~M. Vermaseren, \emph{{The Three-loop splitting
  functions in QCD: The Singlet case}},
  \href{https://doi.org/10.1016/j.nuclphysb.2004.04.024}{\emph{Nucl. Phys.}
  {\bfseries B691} (2004) 129--181},
  [\href{https://arxiv.org/abs/hep-ph/0404111}{{\ttfamily hep-ph/0404111}}].

\bibitem{Bukhvostov:1985rn}
A.~P. Bukhvostov, G.~V. Frolov, L.~N. Lipatov and E.~A. Kuraev,
  \emph{{Evolution Equations for Quasi-Partonic Operators}},
  \href{https://doi.org/10.1016/0550-3213(85)90628-5}{\emph{Nucl. Phys.}
  {\bfseries B258} (1985) 601--646}.

\bibitem{Becchi:1975nq}
C.~Becchi, A.~Rouet and R.~Stora, \emph{{Renormalization of Gauge Theories}},
  \href{https://doi.org/10.1016/0003-4916(76)90156-1}{\emph{Annals Phys.}
  {\bfseries 98} (1976) 287--321}.

\bibitem{Joglekar:1975nu}
S.~D. Joglekar and B.~W. Lee, \emph{{General Theory of Renormalization of Gauge
  Invariant Operators}},
  \href{https://doi.org/10.1016/0003-4916(76)90225-6}{\emph{Annals Phys.}
  {\bfseries 97} (1976) 160}.

\bibitem{Joglekar:1976eb}
S.~D. Joglekar, \emph{{Local Operator Products in Gauge Theories. 1.}},
  \href{https://doi.org/10.1016/0003-4916(77)90014-8}{\emph{Annals Phys.}
  {\bfseries 108} (1977) 233}.

\bibitem{Joglekar:1976pe}
S.~D. Joglekar, \emph{{Local Operator Products in Gauge Theories. 2.}},
  \href{https://doi.org/10.1016/0003-4916(77)90170-1}{\emph{Annals Phys.}
  {\bfseries 109} (1977) 210}.

\bibitem{collins_1984}
J.~C. Collins, \emph{Renormalization}.
\newblock Cambridge Monographs on Mathematical Physics. Cambridge University
  Press, 1984.

\bibitem{Braun:2018mxm}
V.~M. Braun, A.~N. Manashov, S.~Moch and M.~Strohmaier, \emph{{Conformal
  symmetry of QCD in $d$-dimensions}},
  \href{https://arxiv.org/abs/1810.04993}{{\ttfamily 1810.04993}}.

\bibitem{Basso:2006nk}
B.~Basso and G.~P. Korchemsky, \emph{{Anomalous dimensions of high-spin
  operators beyond the leading order}},
  \href{https://doi.org/10.1016/j.nuclphysb.2007.03.044}{\emph{Nucl. Phys.}
  {\bfseries B775} (2007) 1--30},
  [\href{https://arxiv.org/abs/hep-th/0612247}{{\ttfamily hep-th/0612247}}].

\bibitem{Braun:2011dg}
V.~M. Braun and A.~N. Manashov, \emph{{Operator product expansion in QCD in
  off-forward kinematics: Separation of kinematic and dynamical
  contributions}}, \href{https://doi.org/10.1007/JHEP01(2012)085}{\emph{JHEP}
  {\bfseries 01} (2012) 085},
  [\href{https://arxiv.org/abs/1111.6765}{{\ttfamily 1111.6765}}].

\bibitem{Braun:2009mi}
V.~M. Braun, A.~N. Manashov and B.~Pirnay, \emph{{Scale dependence of
  twist-three contributions to single spin asymmetries}},
  \href{https://doi.org/10.1103/PhysRevD.80.114002,
  10.1103/PhysRevD.86.119902}{\emph{Phys. Rev.} {\bfseries D80} (2009) 114002},
  [\href{https://arxiv.org/abs/0909.3410}{{\ttfamily 0909.3410}}].

\bibitem{Spiridonov:1984br}
V.~P. Spiridonov, \emph{{Anomalous Dimension of $G^2_{\mu\nu}$ and $\beta$
  Function}}, {\ttfamily preprint IYaI-P-0378}.

\end{thebibliography}

\providecommand{\href}[2]{#2}\begingroup\raggedright\endgroup


\end{document}